\documentclass[fleqn,usenatbib]{mnras}

\usepackage{newtxtext,newtxmath}
\usepackage[T1]{fontenc}

\DeclareRobustCommand{\VAN}[3]{#2}
\let\VANthebibliography\thebibliography
\def\thebibliography{\DeclareRobustCommand{\VAN}[3]{##3}\VANthebibliography}

\usepackage{graphicx}	
\usepackage{amsmath}	
\usepackage{xcolor}
\usepackage{array} 
\usepackage{orcidlink}

\newcommand{\pctworay}{\texttt{pyC$^2$Ray}}
\newcommand{\mhalo}{M_\mathrm{h}}
\newcommand{\mstar}{M_\star}
\newcommand{\fesc}{f_\mathrm{esc}}
\newcommand{\fstar}{f_\star}
\newcommand{\sfr}{\dot{M}_\star}
\newcommand{\dotN}{\dot{N}_\gamma}
\newcommand{\Muv}{M_\mathrm{UV}}
\newcommand{\Msun}{\,\mathrm{M_\odot}}
\newcommand{\Msunyr}{\,\mathrm{M_\odot/yr}}
\newcommand{\cMpc}{\,\mathrm{cMpc}}
\newcommand{\spice}{\texttt{SPICE}}
\newcommand{\fire}{\texttt{FIRE-2}}
\newcommand{\thesan}{\texttt{Thesan}}

\defcitealias{Hirling2024pyc2ray}{Hirling \& Bianco et al. 2024}
\defcitealias{Bianco2026pasc}{Bianco et. al 2026}

\newcommand{\hii}{\rm{H\scalebox{1.0}[0.8]{II}}}
\newcommand{\hi}{\rm{H\scalebox{1.0}[0.8]{I}}}

\newcolumntype{M}[1]{>{\centering\arraybackslash}m{#1}} 

\title[Consistent EoR]{Consistently Reproducing Early Galaxy Observations and 21-cm Signal during the Epoch of Reionization}

\author[Bianco et al.]{Michele Bianco\orcidlink{0000-0002-6766-0017},$^{1}$\thanks{E-mail: mbianc@ethz.ch}
Sambit K. Giri\orcidlink{0000-0002-2560-536X},$^{2}$
Arghyadeep Basu\orcidlink{0000-0001-8104-9751},$^{3}$
Tommaso Boschi\orcidlink{0000-0002-7226-8255},$^{1}$
Aurel Schneider\orcidlink{0000-0001-7055-8104},$^{4}$ \newauthor
Enrico Garaldi\orcidlink{0000-0002-6021-7020},$^{5,6}$ 
Garrelt Mellema\orcidlink{0000-0002-2512-6748},$^{2}$
\\
$^{1}$Institute for Particle Physics \& Astrophysics (ETHZ), Wolfgang-Pauli-Str 27, 8093 Zurich, Switzerland\\
$^{2}$Department of Astronomy and Oskar Klein Centre, AlbaNova, Stockholm University, SE-10691 Stockholm, Sweden\\
$^{3}$Centre de Recherché Astrophysique de Lyon, Univ Lyon1, Ens de Lyon, CNRS, CRAL UMR5574, F-69230 Saint-Genis-Laval, France\\
$^{4}$University of Zurich, Winterthurerstrasse 190, 8057 Zurich, Switzerland\\
$^{5}$Kavli IPMU (WPI), UTIAS, The University of Tokyo, Kashiwa, Chiba 277-8583, Japan\\
$^6$ Center for Data-Driven Discovery, Kavli IPMU (WPI), UTIAS, The University of Tokyo, Kashiwa, Chiba 277-8583, Japan\\
}

\date{Accepted XXX. Received YYY; in original form ZZZ}

\pubyear{\the\year{}}

\begin{document}
\label{firstpage}
\pagerange{\pageref{firstpage}--\pageref{lastpage}}
\maketitle

\begin{abstract}
In the standard cosmological model, galaxies forming within dark matter haloes are the primary drivers of the Epoch of Reionization (EoR). The complex physics of galaxy formation introduces scatter in the relation between halo mass and star-formation activity, with implications for the production and escape of ionising photons. In this work, we implement a stochastic source model in our cosmological radiative transfer simulations. We calibrate our model against the high-redshift ultraviolet luminosity function observed by JWST and observational constraints on Lyman-continuum leakage to constrain the ionising escape fraction. We find that stochastic star formation redistributes ionising emissivity toward lower-mass haloes, producing a more fragmented ionisation topology and delaying the growth of the largest ionised regions relative to the deterministic, observationally calibrated model. However, when the two models match total ionising photon production, their global neutral fractions and 21-cm power spectrum remain nearly indistinguishable, despite small differences in the morphology of individual ionised regions. Additionally, we establish a unified framework that lets us apply source models derived from computationally expensive hydrodynamical simulations (\spice{}, \fire{} and \thesan{}) and investigate their impact on large-volume reionisation simulations, extending their prescriptions to scales of >100 Mpc. We find that these prescriptions produce distinct ionisation topologies and 21-cm statistical signatures, even when their global reionisation histories are similar. Our results highlight the importance of combining physically motivated galaxy formation prescriptions from hydrodynamical simulations with large-volume radiative transfer simulations to interpret forthcoming 21-cm observations and constrain the sources of cosmic reionisation.

\end{abstract}


\section{Introduction}
Probing the Epoch of Reionization (EoR) is essential for studying the structure formation and evolution of our Universe. During this period, the neutral hydrogen (\hi{}) in the intergalactic medium (IGM) transitions from a predominantly cold and neutral state to a hot and ionised one, due to the appearance of the first luminous cosmic sources \citep[e.g.,][]{furlanetto2006cosmology, Choudhury2006phys, zaroubi2012epoch, Wise2019intro}. These sources, which may have included star-forming galaxies, primordial black holes, X-ray bursts, shock-heated interstellar medium and other secondary sources \citep{Eide2018heat, Eide2020xray, Ma2022crash, Basu2026infl}, produced enough ionising photons to complete reionisation over about 500 million years after the first source formed \citep{haardt2012radiative, Madau2014sfr}.

Radio interferometry telescopes such as the Low Frequency Array \citep[LOFAR;][]{Offringa2012lofar}, the Hydrogen Epoch of Reionization Array experiment \citep[HERA;][]{DeBoer2017hera}, the Murchison Wide-field Array \citep[MWA;][]{trott2020deep} and the Square Kilometre Array Observatory \citep[SKAO;][]{swg2026skao} aim to provide direct observation of the 21-cm signal during the EoR. Some experiments are already providing upper limits \citep[e.g.][]{Patil2017lofar, mertens2020improved, hera2022upper, Mertens2025upper, Nunhokee2025mwa, Abdurashidova2026hera}, but the forthcoming low-frequency component of the SKAO \citep[SKA-Low;][]{mellema2013reionization} will usher us into the era of detection and mapping of the 21-cm signal during reionisation \citep{koopmans2015cosmic, ska2026imaging}. At the same time, infrared space telescopes such as the James Webb Space Telescope \citep[JWST,][]{Gardner2006jwst} and the Nancy Grace Roman Space Telescope \citep[Roman,][]{Spergel2015roman} aim to study the nature and properties of the primary sources of radiation that drive reionisation, with the former already providing some observational evidence. SKA-Low is expected to produce the first data in the coming years, urging the scientific community to better understand the nature of ionising sources and their effect on the neutral IGM. 

Understanding the properties of primordial sources and their interplay with the complex distribution of photon sinks in the IGM may be key to understanding the reionisation process \citep{Finkelstein2019esc, GagnonHartman2025crosscorr}. This is currently done with cosmological hydrodynamical and semi-numerical simulations. Fully numerical radiative hydrodynamical simulations are considered the most physically motivated approach to self-consistently model galaxy formation during the EoR. However, simulating representative cosmological volumes ($\gtrsim100\,\cMpc$), necessary to sample the EoR large-scale variance at $k\leq0.1\mathrm{Mpc}^{-1}$ \citep{iliev2014simulating, bianco2021impact, giri2023suppressing}, and sampling extremely large halos with radiative transfer remain computationally challenging. Therefore, using these simulations for Bayesian inference and large parameter studies of galaxy formation, or for direct comparison with SKA-Low observations, is computationally prohibitive, as they are generally limited to a few realisations \citep[e.g.][]{Ocvirk2016codaI, Ocvirk2020codaII, garaldi2022thesan, Garaldi2024thesan, Zier2026lumina}. Semi-numerical methods have the advantage of being computationally more efficient and able to simulate large volumes \citep[e.g.][]{mesinger201121cmfast, Mutch2016meraxes, Henriques2019lgalaxies, Hutter2021astraeusI, Schaeffer2023beorn, Ma2026polar}. Semi-numerical simulations often post-process dark matter simulations and hence require analytical methods to model the formation of ionising sources. Moreover, they generally suffer from non-conserving photo-production, though some recent work has addressed this limitation \citep[e.g.][]{choudhury2018photon}, and they require assumptions to model the complex interplay between radiation and gas dynamics \citep{Santos2010simfast, Mesinger2016eos, Hutter2018acc, Munoz2022eos}. These limitations motivate the development of simplified yet physically realistic source models for EoR simulations that retain the essential physics of galaxy formation while remaining computationally efficient, a key goal of this paper. 

Numerical and semi-numerical methods rely on dark matter properties to define the source properties \citep[e.g.][]{Davies2025eff}. Meanwhile, several previous works have employed high-resolution hydrodynamical simulations, coupled with radiative transfer, to study the formation and evolution of ionising sources in the high-redshift Universe \citep[e.g.][]{Ceverino2018firstlight, Bird2022astrid, Davies2023comp, Pallottini2023stochsfr, Finkelstein2024ceers}. These studies show that the star formation rate (SFR) in galaxies varies non-linearly, with episodes of intense radiation emission (bursts) followed by periods of stellar inactivity (quiescence) driven by a range of effects. This is the result of photo-evaporation due to radiative feedback \citep[e.g.][]{Trebitsch2017Fluct, Kimm2017Feedback}, supernova explosions \citep[e.g.][]{Kimm2014esc}, anisotropic escape fraction \citep[e.g.][]{Yeh2023thesanfesc}, stellar metallicity variation and different galaxy merger histories \citep[e.g.][]{Paardekooper2015esc}. 

Moreover, the most recent JWST measurement of the ultraviolet luminosity function (UVLF) at high redshift, $z>11$, shows a very high abundance of bright UV sources \citep[e.g.][]{Adams2023jwst, atek2023revealing, Castellano2022jwst, Finkelstein2022jwstuvlf} that could indicate inaccuracies in either the current galaxy formation and/or their astrophysical properties extrapolation \citep{ArrabalHaro2023, Dekel2023effic, Munoz2024crisis, Kokorev2025}. This can be due to either imprecisions in the dust attenuation model \citep{Ferrara2023dustjwst, Markov2024dust}, assumptions on the stellar initial mass function (IMF) in early galaxies \citep{Cueto2024astraeusIX, Trinca2024expl, Ma2025imf}, the accretion and formation of dark matter haloes \citep{Fakhry2026match, Iocco2024compat}, an overly simplistic star formation or feedback model, missing elements in the cosmological model \citep{Menci2022cosmojwst, BoylanKolchin2023stress, Gong2023fuzzy, Jiao2023early, Davari2024spher} or a combination of these effects.

The work by \cite{ren2019brightest} first proposed a semi-empirical model to include scatter in galaxy luminosities, flattening the UVLF at $z>6$. Subsequently, \cite{mason2023brightest} argued that stochasticity in the star-formation rate and the inferred scatter in $\Muv$ could explain the discrepancy in the UVLF at high redshift between models and observations. For this reason, recent work has incorporated stochasticity in the source model of EoR simulations to examine its impact on large-scale 21-cm observables due to this excess of bright sources at $z>13$ \citep[e.g.][]{Furlanetto2022bursty, Hassan2022astro, Bera2025reconciling}. For instance, \cite{Gelli2024} pointed out that the UV absolute magnitude distribution, $\Muv$, measured in the \fire{} simulation \citep{Sun2023Bursty} have a standard deviation inversely proportional to the halo escape velocity, $\sigma_\mathrm{UV}\propto \mhalo^{-1/3} \propto v^{-1}_\mathrm{esc}$, indicating that low-mass haloes are more stochastic at all redshifts. Meanwhile, \cite{Basu2025variabilityuvlf} consider different scenarios of supernova feedback (bursty, smooth, and hyper explosions) with the \spice{} simulations \citep{Bhagwat2024spice}, which induce scatter in the star formation history and the $\Muv$ probability density distribution that depends on both stellar mass and redshift. Therefore, recent work favours a scenario in which primordial sources undergo bursty star formation rather than having extremely high emissivity \citep[e.g.][]{Yung2023, Shen2023variab, Sun2023see, Nikolic2024, Li2024feedback}, which agrees with the stochastic nature implied by early galaxy observations. Moreover, current efforts are aiming to quantify the impact of galaxy stochasticity on the reionisation process \citep[e.g.][]{zhu2026galaxytogalaxystochasticitylimitedimpact}

This paper therefore presents a unified framework that consistently reproduces high-redshift galaxy-formation observables and the 21-cm signal during reionisation. Motivated by recent improved measurements of the Lyman-continuum escape fraction \citep{Giovinazzo2026breaking} and JWST measurements of the UVLF at $z>13$ \citep{bouwens2022z, Castellano2022jwst, Finkelstein2022jwstuvlf, McLeod2024, Kokorev2025glimps, Atek2026glimpse}, we calibrate the free parameters of our stochastic source model directly from observations. We then compare the resulting observationally inferred source models with equivalent prescriptions drawn from three state-of-the-art hydrodynamical simulations, namely \fire{} \citep{Sun2023Bursty}, \spice{} \citep{Basu2025variabilityuvlf}, and \thesan{} \citep{kannan2022introducing}, from which we derive physically motivated source models that we implement in our radiative-transfer code \pctworay{}. This unified framework lets us isolate the impact of stochastic star formation on the timing and morphology of reionisation, assess consistency between observationally inferred and simulation-derived galaxy populations, and investigate how these assumptions propagate into observable 21-cm statistics. 

The paper is organised as follows. In \S\ref{sec:simulations}, we describe the cosmological simulations and radiative-transfer calculations. In \S\ref{sec:source}, we introduce the stellar-to-halo relation, the halo accretion model, and the UV escaping fraction model, which we implemented in our stochastic source model. In \S\ref{sec:results}, we present the observational calibration of the source model, investigate the impact of stochastic star formation on the reionisation process, and compare observationally calibrated source models with those derived from the \fire{}, \spice{}, and \thesan{} hydrodynamical simulations. Finally, in \S\ref{sec:conclusion}, we summarise our main findings and discuss their implications for forthcoming JWST and SKA-Low observations.

\section{Cosmic Reionisation Simulation}\label{sec:simulations}
We describe our methodology for large-scale structure formation and cosmic reionisation in \S\ref{sec:sim_lss} and \S\ref{sec:sim_eor}, respectively. In \S\ref{sec:sim_21cm}, we illustrate how we calculate the 21-cm signal.

\subsection{Simulating Large Scale Structure Formation}\label{sec:sim_lss}
The cosmological structure formation is modelled with the \texttt{Pkdgrav3}\footnote{\url{https://github.com/CLS-project/PKDGRAV3}} code \citep{potter2017pkdgrav3}. We run an N-body simulation with $2048^3$ dark matter particles and a box size of $295\cMpc$ along each direction, resulting in a particle mass of $m_\mathrm{part}=1.2\times10^8\Msun$. We selected this volume size and minimum mass to capture the large-scale fluctuations \citep{iliev2014simulating} and to sample the cosmological large-scale structure relevant to the SKA-Low experiment \citep{giri2023suppressing}. We employ a friends-of-friends algorithm to identify dark matter haloes with at least 10 dark matter particles\footnote{We note that this minimum number of particles used to define a halo is low. Therefore, we validated the halo catalogue used in this work in Appendix A of \citet{Giri2024end}.}, modelling haloes down to a mass of $\sim 10^9\Msun$. This mass resolution implies the assumption that low-mass atomically cooling haloes, with mass $10^8\Msun\leq \mhalo\leq10^9\Msun$, and even smaller haloes are radiatively suppressed and do not host any stars \citep{Dixon2015large, nebrin2023starbursts}. 

We generate initial conditions at $z=150$ using second-order Lagrangian perturbation theory in \texttt{Pkdgrav3} and a transfer function generated with the Boltzmann code \texttt{CLASS} \citep{lesgourgues2011cosmic, blas2011cosmic}. We run the N-body simulation and save the output every $\sim 11\,\mathrm{Myr}$ from $z=90$ to $z=4.55$, yielding $120$ snapshots. From each snapshot, we create mass density fields with a mesh size of $256$, employing a Piecewise Cubic Spline mass assignment scheme \citep{sefusatti2016accurate}, and generating a mesh grid with a spatial resolution of $\sim1.15\cMpc$. We employ a cold dark matter cosmological model constrained by \cite{collaboration2020planck} with parameters $\Omega_\mathrm{m}=0.32$, $\Omega_\mathrm{b}=0.044$, $H_0=67.66\,\mathrm{km\,s^{-1}Mpc^{-1}}$. For the primordial gas, we consider a helium mass-abundance factor of $Y=0.249$ and the corresponding molecular weight $\mu=\left(1-Y\right)^{-1}=1.33$.

\begin{figure*}
    \centering
    \includegraphics[width=0.9\textwidth]{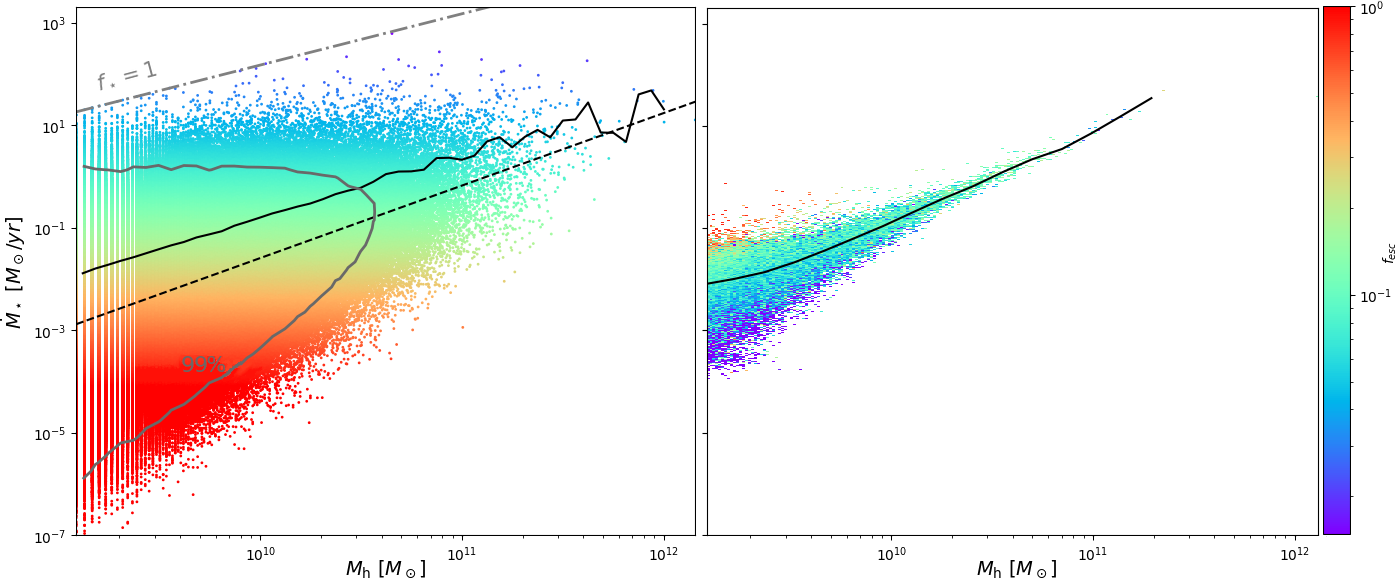}\vskip-2mm
    \caption{Effect of scatter on the relation between halo mass and star-formation rate at $z=9$. \textit{Left Panel}: stochastic $\sfr$ from the observational fit model. A grey dotted-dashed line indicates the physical upper limit in our model, which implies $\fstar = 1$, while the grey lines indicate the $99\%$ contours. The black dashed line is the deterministic relation from the fiducial model for reference. \textit{Right Panel}: The same relation from the \thesan{} simulation data. In both panels, the colorbar indicates the corresponding escaping fraction, while a black solid line indicates the mean of the distribution.}
    \label{fig:scatter_model}
\end{figure*}
\subsection{Simulating the Reionisation of the Intergalactic Medium}\label{sec:sim_eor}
To simulate the propagation of ionising UV radiation, we employ the \pctworay{}\footnote{\url{https://github.com/cosmic-reionization/pyC2Ray}} code \citep{Hirling2024pyc2ray, Bianco2026pasc} to model ionising radiation propagation in post-processing on a cosmological simulation volume. This code is an updated version of the \texttt{C$^2$Ray} code originally developed by \cite{mellema2006c2}, and it includes a Graphics Processing Unit (GPU) based refactoring of the three-dimensional radiative transfer equation for speed-up and a user-friendly \texttt{Python} interface.

In \pctworay{}, we populate dark matter haloes with clusters of primordial sources. An analytical model then quantifies the number of photons emerging from these haloes, of mass $\mhalo$, in the form \citep{ giri2021measuring, Giri2024end}:
\begin{equation} \label{eq:dotN}
    \dotN(\mhalo,z) = N_\mathrm{ion} \,\fstar(\mhalo)\,\fesc(\mhalo)\,\frac{\dot{M}_\mathrm{h}(\mhalo,z)}{(\Omega_b/\Omega_m) \mu m_\mathrm{p}} \ .
\end{equation}
Here $\fstar$, $\fesc$, and $\dot{M}_\mathrm{h}$ are the stellar-to-halo relation, the photon-escape fraction and the halo accretion rate, respectively. While $N_\mathrm{ion}$ is the efficiency of ionising-photon production, $\mu$ is the molecular weight, and $m_\mathrm{p}$ is the proton mass.

This prescription provides a flexible parametrisation of the ionising-source model that drives the reionisation process, and it can be defined by observational constraints on the ionising source or by self-consistent hydrodynamical simulations of galaxy formation and the EoR.

\subsection{Simulating the 21-cm Signal during EoR}\label{sec:sim_21cm}
The presence of neutral hydrogen (\hi{}) during the EoR can be quantified by measuring the intensity of the 21-cm signal emerging from a gas cloud. Radio interferometry telescopes observe this signal in contrast against the CMB temperature, in what is known as the differential brightness \citep[e.g.][]{barkana2001beginning, furlanetto2006cosmology, zaroubi2012epoch, pritchard201221}

\begin{multline}
    \begin{split}
        \delta T_b(\mathbf{n},z) &= \frac{T_\mathrm{S} - T_\mathrm{CMB}}{1+z} \left(1 - e^{-\tau_\nu} \right) \\
        & \simeq 
        T_0\,x_{\rm HI}(1+\delta_b)
        \left(1-\frac{T_\mathrm{CMB}}{T_S}\right)
        \frac{(1+z)^2}{H(z)/H_0} \ ,
    \end{split}
\end{multline}

where $T_S$ is the spin temperature of the gas, $\delta_b$ is the baryonic gas overdensity and $x_\mathrm{HI}$ is the hydrogen neutral fraction. The constant temperature $T_0 $ has numerical value of $\sim 4.72\,\rm mK$. For our simulations, we assume spin-temperature saturation, also known as the heating-saturated approximation, where $T_S \gg T_\mathrm{CMB}$. Previous work shows that the IGM is heated well above the CMB temperature from QSO-like sources and X-ray binary stars in the early stage of reionisation \citep[e.g.][]{ghara201521, ross2017simulating}. This assumption implies that $\delta T_b$ is strictly positive at lower redshift. 

\section{Source \& Sinks Model for Reionisation}\label{sec:source}
In \pctworay{}, we must explicitly specify the photon production rate ($\dotN$) of the ionising sources and the sub-grid sinks of radiation in our simulation volume. In this section, we present the key components of our source model, which was developed in \citet{Giri2024end} to model high-redshift luminosity function data. 

\subsection{Stellar-to-Halo Ratio and Accretion Model}\label{sec:fstar_dotM}
We model the fraction of stellar mass in dark matter haloes, the stellar-to-halo ratio, by a power-law function that depends on the mass of its host $\mhalo$, as given in \citep{park2019inferring, schneider2021halo, schneider2023cosmological},
\begin{equation}\label{eq:fstar}
    \fstar(\mhalo)=f_{\star,0}\,\left(\frac{\mhalo}{10^{10}\Msun}\right)^{\alpha_\star} \ ,
\end{equation}
where $f_{\star,0}$ and $\alpha_\star$ are free parameters. We model the halo accretion rate as an exponential model
that gives \citep{Furlanetto2017minimalist, schneider2021halo}
\begin{equation} \label{eq:acr_rate}
    \dot{M}_\mathrm{h}(z) 
    = \frac{\mhalo}{\left[ \alpha_\mathrm{h}(1+z)\,H(z) \right]^{-1}} \ ,
\end{equation}
where $\alpha_\mathrm{h}$ is a free parameter and $H(z)$ is the redshift dependent Hubble function. With these two equations, we define the star-formation rate (SFR) as $\sfr = \fstar\frac{\Omega_\mathrm{b}}{\Omega_\mathrm{m}}\,\dot{M}_\mathrm{h}$. The UV absolute magnitude ($\Muv$) of these sources can be given by the relation
\begin{equation}\label{eq:Muv}
    \Muv = M_0 - 2.5\,\log_{10}\left(\frac{\sfr}{\kappa}\right) \ ,
\end{equation}
where the two constants are $M_0 = 51.6$, corresponding to an AB magnitude system \citep{Oke1974ABsystem}, and $\kappa = 1.15\times10^{-28}\mathrm{\Msun\,yr^{-1}/(Hz^{-1}s^{-1}erg)}$, calibrated at rest-frame wavelength $1,500$ \AA{} for UV luminosity dust correction, which assumes continuous star formation and a Salpeter stellar initial mass function \citep{Conroy2010calibr}. 

\subsection{Stochastic Star Formation Model }\label{sec:scatter}
A deterministic relation between SFR and $\mhalo$ does not represent star formation measured in several state-of-the-art hydrodynamical simulations, because environmental conditions across haloes of similar masses can affect star formation and evolution. In this paper, we incorporate stochasticity in the SFR by modelling the scatter in the $\Muv-\mhalo$ relation \citep{Gelli2024, Nikolic2024}. We obtain the mean UV magnitude, $\overline{M}_\mathrm{UV}$, from \autoref{eq:Muv} using the one-to-one relation between halo mass and SFR, via \autoref{eq:fstar} and \autoref{eq:acr_rate}. We model the scatter around this relation with a mass-dependent standard deviation given by
\begin{equation}\label{eq:std_Muv}
    \sigma_\mathrm{UV}(\mhalo) = a_s \cdot \log_{10}\left(\frac{\mhalo}{\Msun} \right) + b_s \ ,
\end{equation}
where $a_s<0$ and $b_s>0$ are free parameters. This parametrisation generalises the $\Muv-M_\mathrm{h}$ scatter measured in high-redshift galaxy formation simulations \citep{Sun2023Bursty, Basu2025variabilityuvlf}. For each halo, the UV magnitude is then drawn from a Gaussian distribution, $\Muv \sim \mathcal{N}(\overline{M}_\mathrm{UV},\,\sigma_\mathrm{UV}^2)$ \citep{Sahlen2021}. By inverting each realisation with \autoref{eq:Muv}, we propagate this stochasticity to the SFR. In \S\ref{sec:obs_fit}, we illustrate how we select the values for $a_s$ and $b_s$.

In the left panel of \autoref{fig:scatter_model}, we show a realisation of our stochastic model for $z=9$, illustrating the relation between $\sfr$ and $\mhalo$. The colorbar indicates the UV escaping fraction, discussed further in \S\ref{sec:fesc}, while the solid black line shows the average SFR. Compared with the right panel, which shows the same relation from the \thesan{} hydrodynamical simulation \citep{kannan2022introducing}, we note that the scatter is reduced at matching halo mass compared to our model. The \thesan{} realisation we use does not prescribe a bursty star formation model. Here, the x-axis is limited to the mass resolution of our simulation, while the original simulated data shows a broader scatter at lower mass, $\mhalo\leq10^9\Msun$.

The majority of haloes in the simulation, $\sim99\%$ with $M_\mathrm{halo} \lesssim 3\times10^{10}\Msun$, exhibit a large scatter in star-formation rate, spanning between $\sfr \sim 10^{-6}$ and $1\Msunyr$. At the low end, the broad scatter produces both UV-bright and UV-faint outliers, with the mean (solid black line) approximately $0.5\,\mathrm{dex}$ above the deterministic relation (dashed black line) at all $\mhalo$. In contrast, higher-mass haloes show more modest scatter, and this mass-dependent behaviour enforces the redistribution of ionising-photon emission between haloes, which we discuss in \S\ref{sec:test1}.

\subsection{UV Escape Fraction Model}\label{sec:fesc}
In the case of the escape fraction, we extend the model employed in our previous work \citep{Giri2024end} to include redshift dependence with the form
\begin{equation}\label{eq:fesc}
    \fesc(\mstar,\,z)=f_\mathrm{esc,0}\,\left(\frac{\mstar}{10^{10}\Msun}\right)^{\alpha_\mathrm{esc,M}}\,\left(\frac{1+z}{6}\right)^{\alpha_\mathrm{esc,z}} \ ,
\end{equation}
where $f_\mathrm{esc,0}$, $\alpha_\mathrm{esc,M}$, and $\alpha_\mathrm{esc,z}$ are free parameters that are calibrated to match the desired reionisation model. This formulation is motivated by recent measurements of Lyman-continuum escape fractions from JWST/NIRSpec observations \citep{Giovinazzo2026breaking}, which constrain a broad range of stellar masses and redshifts. We also adopt a similar formulation from a recent study on the redshift dependence of $\fesc$ \citep{Wang2026fesc}. We will discuss this further in \S\ref{sec:obs_fit}. 

Moreover, in contrast to previous studies that parametrise $\fesc$ as a function of halo mass, $\mhalo$, \citep[e.g.][]{park2019inferring, schneider2023cosmological, Schaeffer2023beorn}, we adopt a stellar-mass-dependent formulation, $\mstar$. This choice is required to consistently incorporate the stochasticity introduced in \S\ref{sec:scatter}, where the mapping between halo mass and star-formation rate—and hence stellar mass—is no longer deterministic. 

We show an example of our model at $z=9$, indicated by the colorbar of \autoref{fig:scatter_model}. Our escape fraction model shows an inverse trend to what is measured in the \thesan{} hydrodynamical simulation. In our case, haloes with high SFR have high $\fesc$, while low-SFR haloes have lower values ($\fesc\leq0.2$). Meanwhile, in \thesan{} (right panel), haloes below/above the average SFR tend to have a low/high escape fraction. In the hydrodynamical simulation, this quantity is measured as the fraction of ionising photons that reach the virial radius relative to the total number produced in each dark matter halo \citep[for more details, see][]{Yeh2023thesanfesc}. Star particles formed outside the virial radius have a fixed value, $\fesc=0.37$. We discuss the implications of these differences in \S\ref{sec:obs_fit}.

\subsection{Sinks of Radiation Model}\label{sec:sinks}
Currently, \pctworay{} includes two models to account for unresolved sinks of UV radiation: a redshift- and density-dependent clumping factor, $\mathcal{C}$ \citep{mao2020impact, bianco2021impact} and a redshift-dependent mean-free path of the ionising photons, $R_\mathrm{mfp}$. In this paper, we consider only the latter and set the clumping factor to a constant, $\mathcal{C}=1$. We model the $R_\mathrm{mfp}$ based on an extended version of the model presented by \cite{worseck2014giant}, which in our case is defined as
\begin{equation}\label{eq:rmfp}
    R_\mathrm{mfp}(z) = A_\mathrm{mfp}\left(\frac{1+z}{5}\right)^{\eta_{\mathrm{mfp},0}}\left[1+\left(\frac{1+z}{1+z_\mathrm{mfp}}\right)^{\eta_{\mathrm{mfp},1}}\right]^{\eta_{\mathrm{mfp},2}} \ ,
\end{equation}
here, we set the four free parameters to be $A_\mathrm{mfp}=210\,\mathrm{cMpc}$, $\eta_{\mathrm{mfp},0}=-9$, $\eta_{\mathrm{mfp},1}=9$, $\eta_{\mathrm{mfp},2}=1$ and $z_\mathrm{mfp}=6$. This parameter choice corresponds to the \textit{SinkA} model from our previous work \citep[see][for more details]{Giri2024end}.

This choice yields an $R_\mathrm{mfp}$ that remains constant for most of the reionisation history, close to $\sim 10\,\mathrm{cMpc}$ for $z \gtrsim 8$. At lower redshift, it grows exponentially to $\sim 50\,\mathrm{cMpc}$ at $z=5$, matching observational measurements of the mean-free path.


\begin{figure}
    \centering
    \includegraphics[width=\columnwidth]{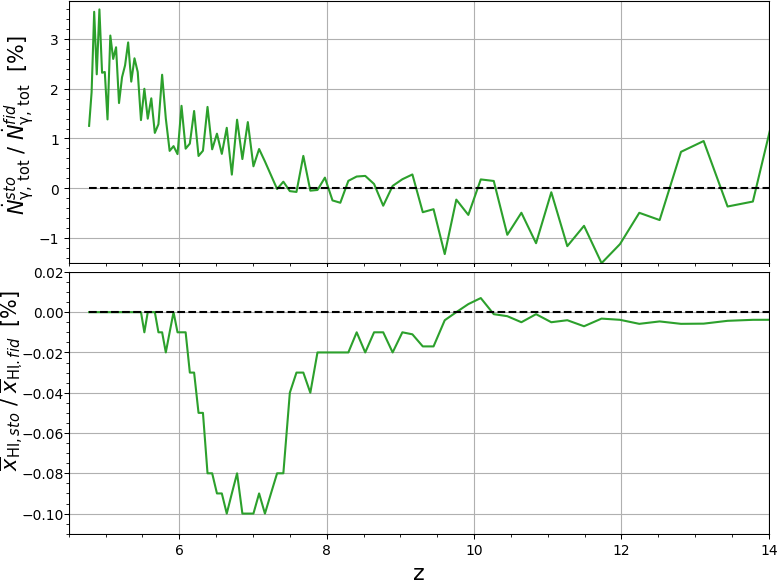}\vskip-2mm
    \caption{Effect of the stochastic star formation model (green solid line) on reionisation against the fiducial model (black dashed line). \textit{Top Panel}: comparison of the grand total production of ionising photons $N_\gamma^\mathrm{tot}$. \textit{Bottom Panel}: corresponding difference in the reionisation history.}
    \label{fig:eor_history_test1}
\end{figure}

\section{Results} \label{sec:results}
Here we present a suite of reionisation simulations using the stochastic source model described in \S\ref{sec:scatter}. In \S\ref{sec:test1}, we assess how scatter in $\sfr$ affects reionisation. In \S\ref{sec:obs_fit}, we calibrate the free parameters using observational constraints on high-redshift galaxies, and in \S\ref{sec:test2}, we calibrate the source model using galaxy formation simulations.

\subsection{Conserving the grand total number of ionising photons}\label{sec:test1}
We first compare the fiducial simulation, which is characterised by a deterministic source model, with a model that incorporates stochastic star formation. For this simulation, we set the source parameters to $f_{\star,0} = 0.1$, $\alpha_\star = -0.30$, $\alpha_\mathrm{h}=0.79$, and a constant escaping fraction with $f_\mathrm{esc}=0.02$. We motivate this choice by our previous work \cite{Giri2024end}, in which we calibrated the source model to capture the 21-cm signal transition from EoR to Post-EoR. 

To ensure a controlled comparison, we tune the free parameters of the stochastic model, $f_{\star,0}$, $\alpha_\star$, $\alpha_\mathrm{h}$, $a_s$ and $b_s$, from \autoref{eq:fstar}, \ref{eq:acr_rate} and \ref{eq:std_Muv}, such that it reproduces the same cumulative ionising photon production as the fiducial case. We therefore calculate the emission rate of ionising photons per unit volume, $\dot{N}_\mathrm{\gamma,tot}(z) = \frac{1}{V_\mathrm{box}} \sum_{i} \dot{N}_\gamma(M_\mathrm{h, i},z)$, of all the ionising sources present at each redshift snapshot. This setup allows us to isolate the impact of stochasticity in the source model on the reionisation history and its statistical properties. 

\begin{figure}
    \centering
    \includegraphics[width=\columnwidth]{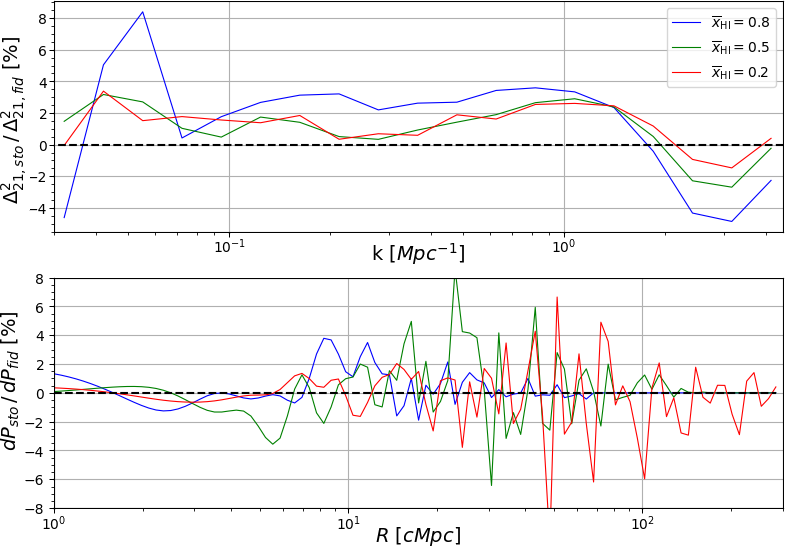}\vskip-2mm
    \caption{Effect of the stochastic model against the fiducial model. \textit{Top Panel}: differences in the 21-cm dimensionless power spectra and \textit{Bottom Panel}: differences in the bubble size distribution at matching reionisation milestones, $\overline{x}_\mathrm{HI} = 0.8$, $0.5$ and $0.2$ indicated by the colours.}
    \label{fig:stat_compare_test1}
\end{figure}

\begin{figure*}
    \centering
    \includegraphics[width=0.95\textwidth]{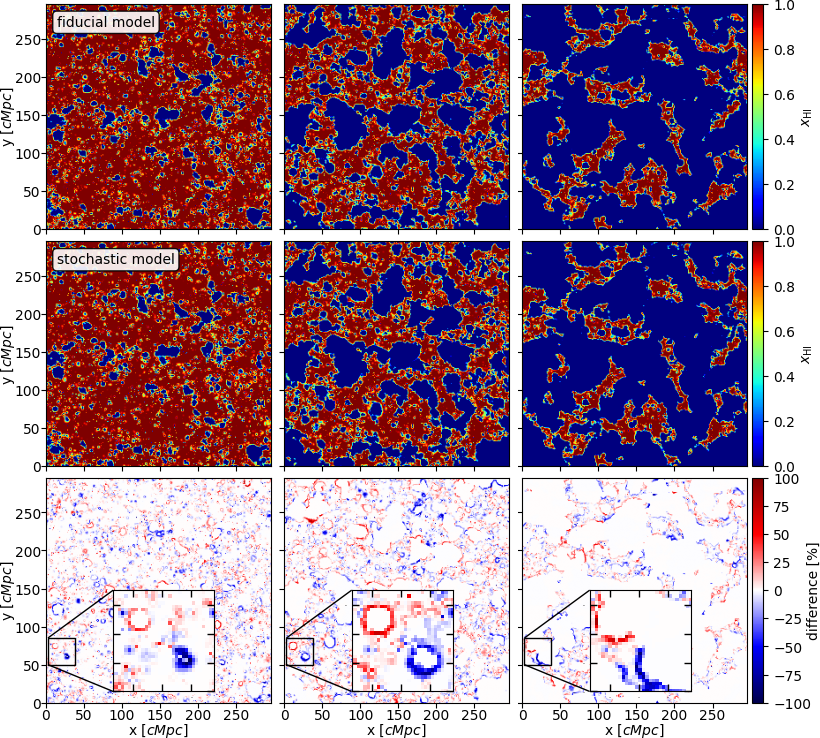}\vskip-3mm
    \caption{Visual comparison of the hydrogen neutral fraction distribution. Comparison between the deterministic (top row) and stochastic source model (middle row) at matching reionisation milestones of $\overline{x}_\mathrm{HI} = 0.8$ (left panels), $0.5$ (middle panels), and $0.2$ (right panels). The bottom panels show the percentage difference between the two maps, with dark blue/red colour indicating regions ionised more in the stochastic/fiducial scenario.}
    \label{fig:xHI_compare_test1}
\end{figure*}

The top panel of \autoref{fig:eor_history_test1} shows the resulting difference in the emission rate of ionising photons per unit volume between the stochastic model (green solid line) and the fiducial model (black dashed line). We note that a difference in $\dot{N}_\mathrm{\gamma,tot}$ within $\pm 3.5\%$ reproduces almost matching global reionisation histories (bottom panel), yielding a maximum difference in the global neutral fraction of $\overline{x}_\mathrm{HI,sto}\,/\,\overline{x}_\mathrm{HI,fid}\sim -0.1\%$. We notice that the difference in $\dot{N}_\mathrm{\gamma,tot}$ in the stochastic model increases at lower redshift, $z<5$. However, since the gas in the simulation is nearly completely ionised, the difference in $\overline{x}_\mathrm{HI}$ is near zero. 

We calculate the 21-cm dimensionless power spectrum, $\Delta^{2}_\mathrm{21}$, and the Bubble Size Distribution (BSD) using the mean-free-path method \citep{mesinger2007efficient, giri2018bubble}. Both were produced using the \texttt{tools21cm} Python package for EoR simulation analysis \citep{giri2020tools21cm}. The top panel of \autoref{fig:stat_compare_test1} shows the difference between the stochastic and fiducial dimensionless 21-cm power spectrum, where line colour denotes the reionisation stage. Previous work has shown how stochasticity in X-ray emission affects the 21-cm signal during EoR \citep[e.g][]{Dasgupta2026impactstochasticpopiiixray}. In contrast, we found that stochasticity in the SFR produces small variations in the global neutral fraction that are not enough to produce evident differences in the dimensionless power spectra, $\Delta^2_\mathrm{21,\,sto}/\Delta^2_\mathrm{21,\,fid}\sim \pm8\%$. Therefore, we expect that the ionising source effects cannot be determined from the global neutral fraction and 21-cm power-spectrum statistics alone, which agrees with the findings of \cite{Davies2025eff}. The bottom panel of \autoref{fig:stat_compare_test1} shows the corresponding difference in the BSD. For this analysis, we used a threshold of $x^\mathrm{thr}_\mathrm{HII}=0.9$ to highlight the highly ionised bubbles. We find that despite differences in the source model, the two realisations produce very similar distributions of ionised regions, making them virtually indistinguishable, with absolute differences within $\sim 8 \%$. In both cases, the average bubble size, defined as $\overline{R} = \int \left(R\,dN/dR \right) dR$, evolves from $\overline{R} = 3.26\cMpc$ (fiducial) and $3.31$ (stochastic) at an early stage ($\overline{x}_\mathrm{HI}=0.8$), to $11.09\cMpc$ and $11.15\cMpc$ at middle ($\overline{x}_\mathrm{HI}=0.5$) and to $30.95\cMpc$ and $30.90\cMpc$ at the end stage of EoR ($\overline{x}_\mathrm{HI}=0.2$).

\begin{figure*}
    \centering
    \includegraphics[width=\textwidth]{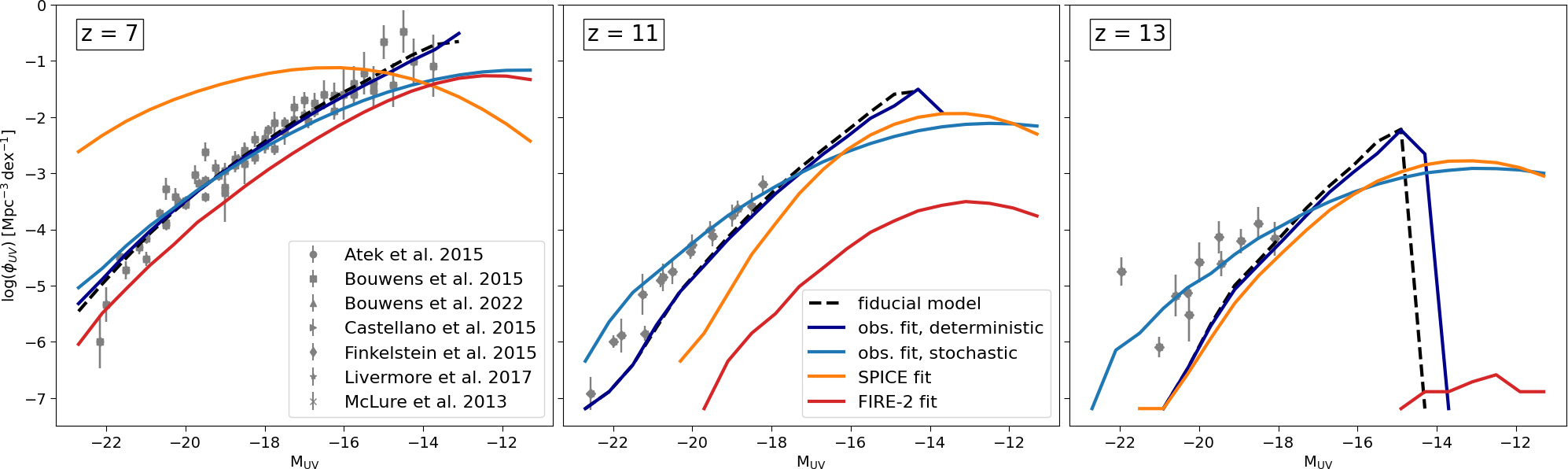}\vskip-3mm
    \caption{UV luminosity function comparison for our different source models at $z=7$, $11$, and $13$. The \thesan{} fit is excluded as we parametrise $\dotN$ directly.}
    \label{fig:uvlf}
\end{figure*}

\autoref{fig:xHI_compare_test1} shows a visual comparison of the ionisation fields. The first and second rows show slices of the neutral hydrogen fraction from the fiducial model and stochastic model, respectively, at three representative stages: an early phase with $\overline{x}_\mathrm{HI}=0.8$ (left column), the midpoint of reionisation (middle column) and the end stage (right column), $\overline{x}_\mathrm{HI} = 0.2$. Blue regions correspond to ionised gas ($x_\mathrm{HI} \approx 0$), while red regions indicate neutral gas ($x_\mathrm{HI} \approx 1$). The third row shows the relative per cent difference between the models, $x_\mathrm{HI, fid}/x_\mathrm{HI, sto}$, where dark red indicates regions more ionised in the fiducial model and dark blue in the stochastic model. Small fluctuations in the voxel values appear at the edge of ionised bubbles, and the vast neutral IGM appears to alter the growth and size of some large-scale ionised regions. This behaviour stems from the large scatter in the star formation rate of low-mass haloes shown in \autoref{fig:scatter_model}, which affects the bubbles randomly, as expected. For instance, in the bottom panel, we zoom in on the region at coordinates $x\in[0,40]\cMpc$ and $y\in[50,90]\cMpc$. Here, we can see two growing ionised regions, one slightly larger in the deterministic case (dark red) and the other larger in the stochastic case (dark blue). By $\overline{x}\simeq0.2$, the two bubbles merge into a larger ionised region that expands to cover the entire zoom-in panel. 

We run several tens of simulations with the same scatter parameters but different random seeds. Each realisation of the stochastic model has a slightly different neutral hydrogen distribution around the ionised regions, while maintaining the same summary statistics mentioned above. However, despite the fact that the stochastic model is not connected to the halo evolution history, we observe that if a region hosts a quenched galaxy or bursty star-formation event at high redshift, producing a photon under- or overabundance due to scattering, its imprint remains visible at later times, as shown in the inset zoom-in panel.

\begin{figure*}
    \centering
    \includegraphics[width=\textwidth]{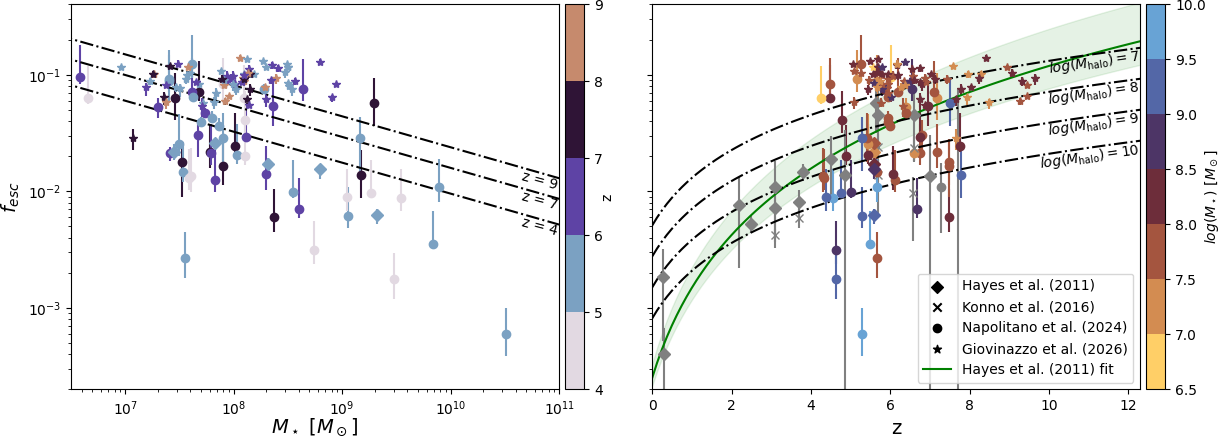}\vskip-3mm
    \caption{Visualisation of our best fit to the redshift- and stellar mass-dependent escape fraction model and the corresponding observational data. \textit{Left Panel}: stellar mass evolution at different redshifts indicated by the colorbar. Black dash-dotted lines show the $\fesc$ model at three fixed redshifts, $z=9$, $7$ and $4$. \textit{Right Panel}: redshift evolution for different stellar masses indicated by the colorbar. Black dash-dotted lines show the $\fesc$ model at $\log(\mstar/\Msun)=7,\,8,\,9$ and $10$. Grey error-bar points represent data without stellar mass information, and the green line indicates previous work fitted to these data.}
    \label{fig:obsfesc}
\end{figure*}

\subsection{EoR source model constrained by observations} \label{sec:obs_fit}
Here, we use observational constraints on high-redshift galaxies to calibrate the free parameters of our source model (\autoref{eq:dotN}). This approach links the observed galaxy population to the ionising emissivity that ionises the IGM. 

The ultraviolet luminosity function (UVLF) directly constrains the star-formation activity of UV-emitting sources and is therefore used to calibrate the free parameters related to the star-formation rate: $f_{\star,0}$, $\alpha_\star$, and $\alpha_\mathrm{h}$. In addition, we include recent UVLF measurements at $z>13$ to constrain the stochasticity parameters $a_s$ and $b_s$ \citep{Castellano2022jwst, Finkelstein2022jwstuvlf, McLeod2024, Kokorev2025glimps}. These high-redshift observations matter because stochasticity primarily affects the abundance of rare, UV-bright galaxies that dominate the bright end of the UVLF.

To constrain the escape fraction, we adopt an empirical relation between Lyman-$\alpha$ and ionising photon escape fractions, $\fesc \approx 0.15\cdot f^{\mathrm{Ly}\alpha}_\mathrm{esc}$, motivated by the observed correlation between these quantities \citep{chisholm2018accurately, Begley2024connecting}. We then constrain the parameters $f_\mathrm{esc,0}$, $\alpha_\mathrm{esc,\, M}$, and $\alpha_\mathrm{esc,\, z}$, using recent measurements of Lyman-continuum leakage that extend both the stellar mass range and redshift coverage of previous samples \citep{Napolitano2024Lya, Giovinazzo2026breaking}.

\subsubsection{Bayesian Inference of the Astrophysical Parameters}
We perform two separate Bayesian inferences using a Markov Chain Monte Carlo (MCMC) framework to constrain the free parameters of the star-formation rate and escape fraction model. In both cases, we assume independent measurements and adopt a Gaussian likelihood with a diagonal covariance matrix. We assume a flat prior and sample the parameter space with the Latin Hypercube Sampling \citep{McKay1979lhs} method in the ranges $\log(f_\mathrm{\star,0}) \in [-3, 0]$, $\alpha_\star \in [-1, 1]$, $\alpha_\mathrm{h} \in [0.5, 2]$ for the deterministic observational fit, while for the stochastic case, we additionally have $a_s\in[-\frac{1}{3}, 0]$ and $b_s\in[-5, 5]$. Moreover, in the deterministic case, inference fails to converge when we include UVLF data at low and high redshifts simultaneously. Consequently, we limited the calibration of the deterministic observational fit to observational data at $z<11$. For the escape fraction model, we set $\log(f_\mathrm{esc,0}) \in [-4, 0]$, $\alpha_\mathrm{esc,M} \in [-2, 1]$ and $\alpha_\mathrm{esc,z} \in [-2, 5]$. We use the affine-invariant ensemble sampler implemented in the \texttt{emcee} \citep{ForemanMackey2013emcee} package.

In \autoref{fig:uvlf}, we show the resulting UVLF fit at six representative redshifts, $z=13.5,\,11,\,10,\,9,\,8,\,7$ and $6$. Grey error bars indicate the observational data employed for the Bayesian analysis \citep{McLure2013, bouwens2015uv, Finkelstein2015, livermore2017directly, atek2018extreme, bouwens2022z, Castellano2022jwst}. As expected, the deterministic observational fit (solid dark blue) struggles to simultaneously reproduce the abundance of bright galaxies at high redshift and the UVLF at lower redshift, similar to our fiducial model. In contrast, the stochastic observational fit (solid blue) successfully reproduces both regimes. As shown in previous studies, stochasticity in the star-formation rate broadens the UV luminosity distribution and naturally flattens the bright end of the UVLF \citep[e.g.][]{Nikolic2024}. In the same plot, we include additional models that we discuss further in \S\ref{sec:test2}.

In \autoref{fig:obsfesc}, we show the inferred escape-fraction model alongside the observational data used for the Bayesian analysis \citep{Hayes2011, Konno2016, Napolitano2024Lya, Giovinazzo2026breaking}. Grey points correspond to measurements without stellar-mass information. The left panel shows the stellar-mass dependence of the escape fraction, with colour indicating redshift, while the right panel shows the redshift evolution for different stellar masses. Dash-dotted curves indicate our best-fitting model evaluated at representative redshifts and stellar masses. For comparison, we also show the model of \cite{Hayes2011} (green line and shaded region), which includes only the redshift evolution. 

\begin{figure*}
    \centering
    \includegraphics[width=0.9\textwidth]{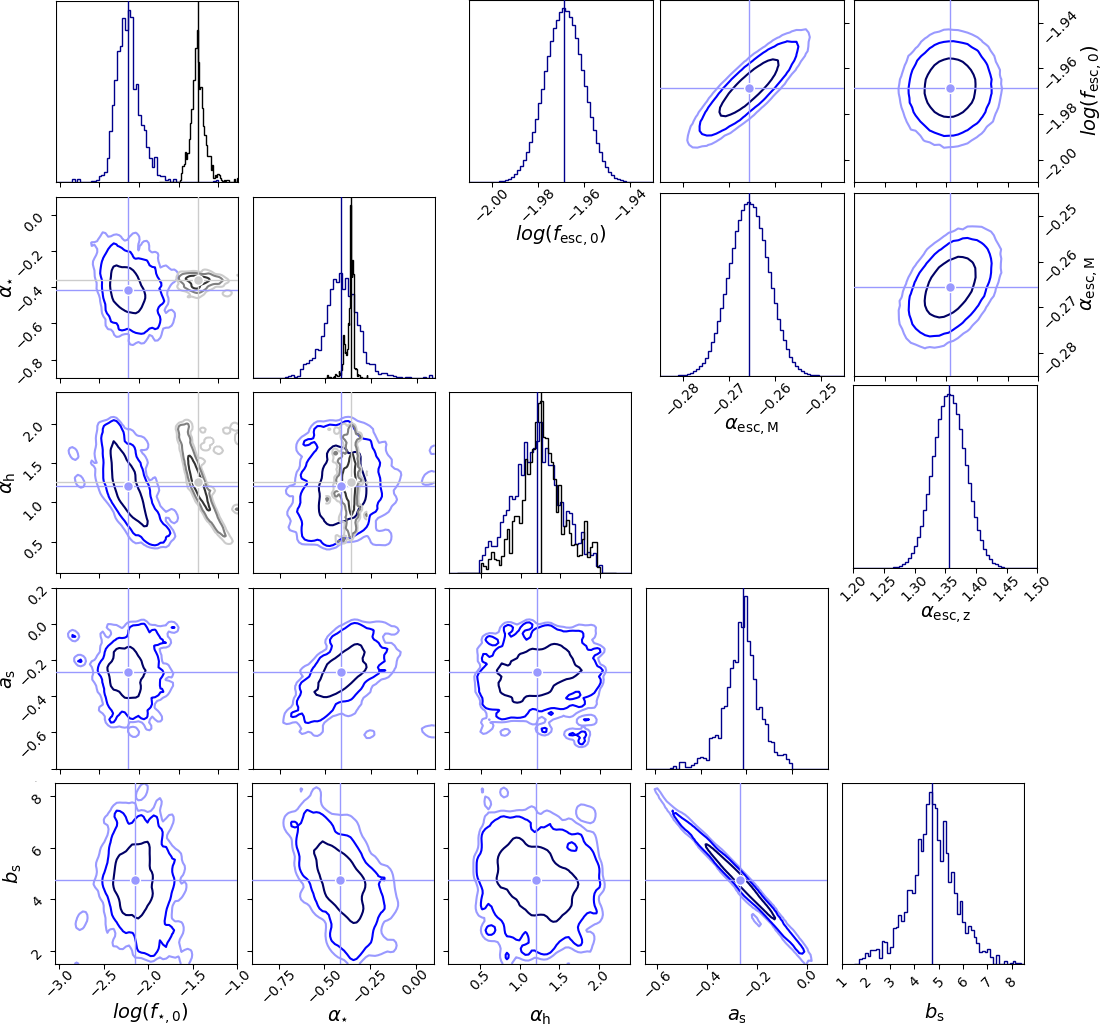}\vskip-2mm
    \caption{Posteriors for the observational source model fit. \textit{Bottom corner}: Posteriors of the five parameters that constrains $\sfr$. Blue/grey contours indicate the stochastic/deterministic observational fit. \textit{Top corner}: Posteriors of the three parameters that constrains $\fesc$. From darker to lighter shades of the same colour indicate the $1\sigma$ ($68\%$), $2\sigma$ ($95\%$), and $3\sigma$ ($99\%$) confidence intervals.}
    \label{fig:corner_plot}
\end{figure*}

In \autoref{fig:corner_plot}, the bottom-left corner shows the posterior on the $\sfr$ parameters, with dark blue contours for the stochastic scenario and grey for the deterministic scenario. Including stochasticity substantially alters the inferred stellar-to-halo relation. In particular, the normalisation parameter $f_{\star,0}$ is about one order of magnitude lower than in the deterministic case. At the same time, the power-law slope varies from $\alpha_\star=-0.361$ in the deterministic fit to $-0.413$. This parameter combination reduces the mass dependence of the star-formation rate and implies that haloes below the pivotal mass of $10^{10}\Msun$ form stars slightly more efficiently in the stochastic scenario. As a result, the mean star-formation efficiency, $\langle\fstar\rangle$, remains close to the one-to-one relation recovered in the deterministic observational fit model. \autoref{tab:bestfit} summarises the best-fit parameters.


In both observational fits, we obtain similar values for $\alpha_\mathrm{h}$, which are significantly larger than the value adopted in our fiducial model. We chose the fiducial value from the exponential accretion-rate model in \cite{Dekel2013toymodel}. Our inferred parameters instead imply that, for the reionisation process, the halo accretion rate is substantially faster, with values approximately $\sfr^\mathrm{obs\,fit}/ \sfr^\mathrm{fiducial}\sim 1.6$ times higher at all redshifts. We attribute this difference to calibrating the fiducial model on dark matter halo-only simulations. This demonstrates that our source model, constrained by galaxy observations, requires faster gas accretion than expected to sustain stellar production. At the same time, both hydrodynamical simulations and observations indicate that the thermal state of the infalling gas and radiative feedback within the host galaxy can significantly delay star formation, particularly in bursty systems \citep[e.g.,][]{Behroozi2015, RodriguezPuebla2017, VanDonkelaar2025phoebos, Sultan2026coldvshotgas}. Future work could replace the exponential prescription with accretion histories derived directly from dark matter halo merger trees \cite[e.g.,][]{Hutter2021astraeusI, Hutter2021astraeusIII, Legrand2021astraeusIV}. As expected, $\alpha_\mathrm{h}$ is strongly degenerate with $f_{\star,0}$, reflecting the coupling between gas accretion from the IGM and the stellar-to-halo mass relation that regulates star formation.

Our posterior favours escape fractions that decrease with increasing stellar mass and decreasing redshift. The top-right panel of \autoref{fig:corner_plot} shows the distributions of the $f_\mathrm{esc}$ parameters. In particular, both the redshift power-law index, $\alpha_\mathrm{esc,z}$, and the normalisation factor, $f_\mathrm{esc,0}$, differ substantially from the values inferred by \cite{Hayes2011}, yielding $\alpha_\mathrm{esc,z}=2.57^{+0.19}_{-0.12}$ and $f_\mathrm{esc,0}=\left(1.67^{+0.53}_{-0.24}\right)\times10^{-3}$.

In \autoref{tab:bestfit}, we list the best-fit parameters from our Bayesian inference for the deterministic and stochastic observational fit scenarios for the SFR model, as well as the parameters for the stellar mass- and redshift-dependent escaping fraction model.
\begin{table}
    \centering
    \caption{Source parameters and best-fit values with relative errors ($1\sigma$) based on galaxy observational constraints for the deterministic and stochastic star-formation scenarios, observational fit.}
    \begin{tabular}{M{1.5cm}|M{2.5cm}M{2.5cm}}\hline
       Parameter            & Deterministic    & Stochastic      \\ \hline\hline \vskip2mm
       $f_{\star,0}$      & $(5.57^{+2.30}_{-1.09})\times 10^{-2}$ & $(7.17^{+2.63}_{-2.27})\times10^{-3}$   \\ \vskip2mm
       $\alpha_\star$       & $-0.361^{+0.016}_{-0.033}$    & $-0.413^{+0.089}_{-0.083}$   \\ \vskip2mm
       $\alpha_\mathrm{h}$       & $1.260^{+0.288}_{-0.241}$     & $1.206^{+0.323}_{-0.317}$    \\ \vskip2mm
       $a_s$                & $-$                           & $-0.268^{+0.086}_{-0.096}$  \\ \vskip2mm
       $b_s$                & $-$                           & $4.731^{+1.201}_{-1.343}$    \\ \hline 
    \end{tabular}
    \begin{tabular}{M{1.6cm}|M{5.4cm}}\hline\vskip1mm
       $f_\mathrm{esc,0}$      & $(1.08\pm0.02)\times10^{-2}$  \\ 
       $\alpha_\mathrm{esc,M}$ & $-0.266\pm0.004$              \\
       $\alpha_\mathrm{esc,z}$ & $1.356\pm0.027$               \\\hline
    \end{tabular}
    \label{tab:bestfit}
\end{table}

\subsubsection{EoR Simulation Constrained by Galaxy Observations}
We use the best-fit parameters listed in \autoref{tab:bestfit} to define two source models and run two EoR simulations with \texttt{pyC$^2$Ray}, which we hereafter refer to as the deterministic and stochastic observational fits. 

\begin{figure*}
    \centering
    \includegraphics[width=0.98\textwidth]{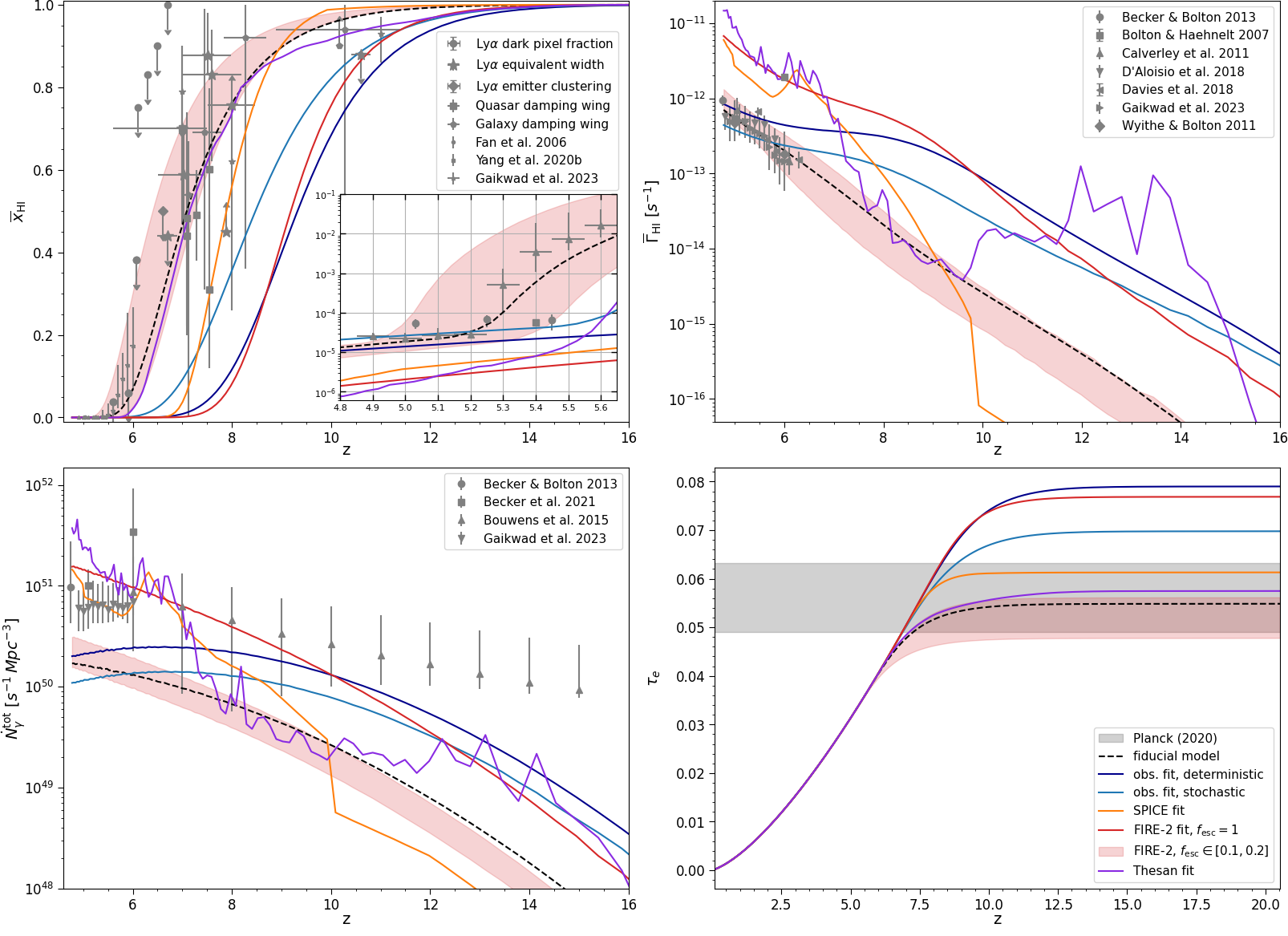}\vskip-3mm
    \caption{Comparison of different source models on the reionisation observational constraints. \textit{Top Left Panel}: redshift evolution of the global average neutral fraction with a zoom-in panel at lower redshift, $z<5.6$. \textit{Top Right Panel}: evolution of the photoionisation rate over redshift. \textit{Bottom Left}: redshift evolution of the ionising photon emission rate per unit volume. \textit{Bottom right}: electron scattering optical depth. Each panel includes observational measurements.}
    \label{fig:eor_obs_constraint}
\end{figure*}

In \autoref{fig:eor_obs_constraint}, we show their effect on the reionisation history and other related EoR observational quantities. In the top-left panel, we show the resulting reionisation histories. An inset panel shows a closer look at the end stages of reionisation ($z \leq 5.6$). Contrary to expectations, and despite similar accretion rates and mean star-formation efficiencies, the two models produce markedly different reionisation histories. We find that the deterministic model ionises faster than the stochastic case. This behaviour arises from the stellar-mass-dependent escape fraction introduced in \S\ref{sec:fesc}. With scatter in $\fstar$, stochastic star formation broadens the $\fesc$ distribution. Therefore, the mean escape fraction in the stochastic model is about a factor of two lower than in the deterministic case at all redshifts. This decrease in $\fesc$ is evident in the global ionising photon emission rate in \autoref{fig:eor_obs_constraint}, bottom-left panel, which shows a constant factor-of-two difference over redshift between the stochastic and deterministic observational fits.

More importantly, both source models calibrated to source observables ionise the Universe substantially earlier than the fiducial model. From the zoom-in panel, both models reach a nearly steady state at the end of reionisation, $z\approx 5$, consistent with the indirect EoR observational inference at low redshift from \cite{gaikwad2023measuring}. However, only the stochastic observational fit is marginally more consistent and agrees better with the measurements, indicating an approximately constant neutral fraction of $\overline{x}_\mathrm{HI} \simeq 2.5 \times 10^{-5}$ at $z<6$. However, both models fail to reproduce the rapid evolution inferred over $5.2 < z < 5.6$ \citep{fan2006constraining, yang2020measurements, gaikwad2023measuring}, which was used to calibrate the fiducial model.

The top-right panel in \autoref{fig:eor_obs_constraint} shows the redshift evolution of the photoionisation rate. Both models show higher rates at higher redshifts, which flatten at $z\sim 8$ for the deterministic model and $z\sim 7$ for the stochastic model. The same trend is evident in the emission rate of ionising photons per unit volume (bottom-left panel), an indicator of the intrinsic luminosity of the ionising sources. The combined effect of a decreasing escape fraction at lower redshift and higher stellar mass explains the decrease in the photon emissivity rate and the photoionisation rate. However, in both cases, the photoionisation rate rises again at $z<6$ due to the increasing mean-free path of the ionising photons, \autoref{eq:rmfp}, which stays constant at approximately $R_\mathrm{mfp}\simeq 10\cMpc$ until $z \sim 8$ and then exponentially increases, allowing the photoionisation rate to once again increase toward the end of EoR.

Free electrons formed during reionisation can interact with CMB photons via inverse Compton scattering, thereby suppressing CMB anisotropies. We quantify this effect with the integrated Thomson scattering optical depth, $\tau_e$. In our case, the bottom-right panel shows that the $\tau_e$ in both our models is well above current observations. This agrees with the conclusion from the top-left panel of \autoref{fig:eor_obs_constraint}, which shows that the deterministic and stochastic models have a faster reionisation history, implying that ionising sources must have begun contributing to reionisation at a higher redshift and over an extended period rather than instantaneously. 

\begin{figure*}
    \centering
    \includegraphics[width=0.83\textwidth]{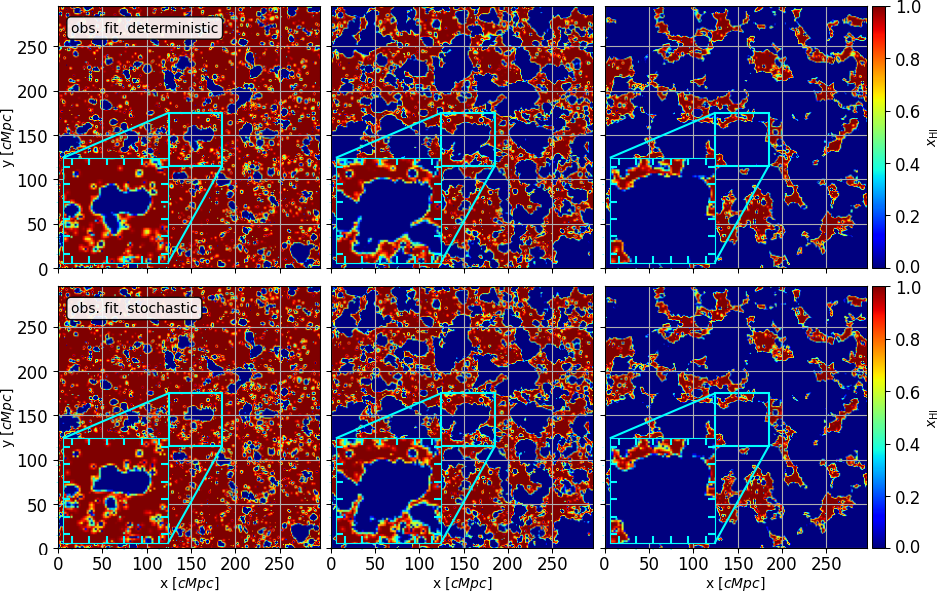}\vskip-3mm
    \caption{Visual comparison of the hydrogen neutral fraction distribution at matching reionisation milestones of $\overline{x}_\mathrm{HI} = 0.8$ (left column), $0.5$ (middle column) and $0.2$ (right column) for the deterministic (top row) and stochastic observational fit (bottom row). An inset zoom-in with a $60\cMpc$ side length shows the evolution of one large ionised region across the different source models.}
    \label{fig:xHI_compare_obsfit}
\end{figure*}

\begin{figure*}
    \centering
    \includegraphics[width=0.9\textwidth]{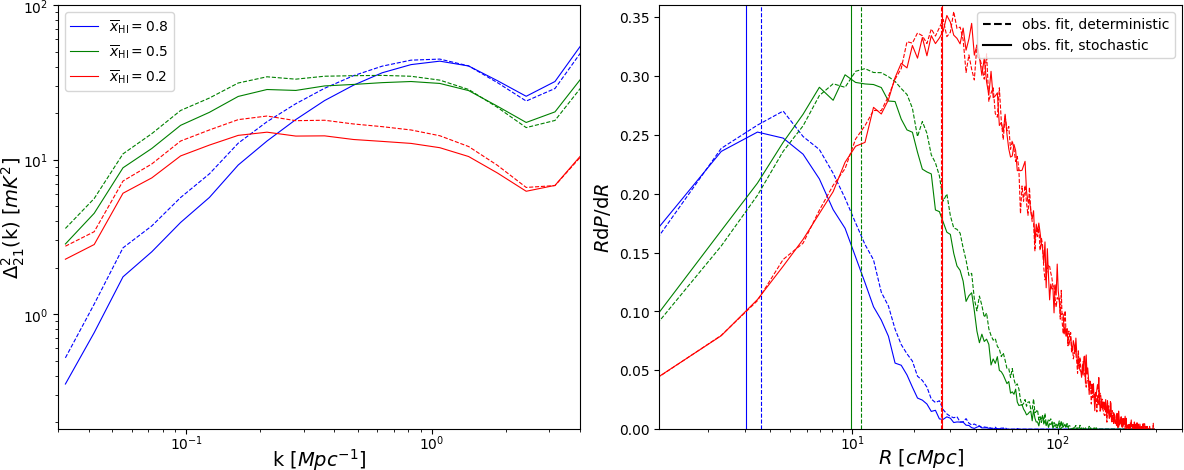}\vskip-3mm
    \caption{Statistical comparison between the deterministic and stochastic observational fit at matching reionisation milestones of $\overline{x}_\mathrm{HI} = 0.8$, $0.5$ and $0.2$. \textit{Left Panel} 21-cm dimensional power spectra. \textit{Right Panel}: bubble size distribution.}
    \label{fig:stat_compare_obsfit}
\end{figure*}

In \autoref{fig:xHI_compare_obsfit}, we show slices of the neutral hydrogen field at the matched reionisation milestones $\overline{x}_\mathrm{HI}=0.8$, $0.5$, and $0.2$. An inset panel zooms in on a $60\cMpc$ region. The first/second rows correspond to the deterministic/stochastic observational fit models. The deterministic observational fit produces slightly larger ionised regions of a few $\cMpc$ in size that, in some cases, merge earlier with nearby \hii{} bubbles. This is particularly evident during the intermediate stages of reionisation, as the zoom-in region shows a larger interconnected ionised region. At the same time, numerous smaller ionised regions emerge earlier in the stochastic case in the predominantly vast neutral IGM.

We perform the same statistical analysis as in \S\ref{sec:test1}. In \autoref{fig:stat_compare_obsfit} left panel, we show the dimensionless 21-cm power spectrum at matched reionisation milestones. The deterministic observational fit exhibits enhanced power on large scales, $k\lesssim1\,\mathrm{Mpc^{-1}}$, relative to the stochastic case throughout the reionisation process. This excess large-scale power reflects the more coherent ionisation field produced by the deterministic source population. In contrast, stochastic star formation promotes smaller ionised regions throughout the neutral IGM, reducing large-scale contrast and suppressing the growth of the largest coherent structures. The bubble size distribution in \autoref{fig:stat_compare_obsfit}, right panel, supports this interpretation. The deterministic scenario consistently exhibits larger bubble sizes for $\overline{x}_\mathrm{HI}>0.3$. This is also evident from the vertical line, which indicates the mean bubble size, $\overline{R}$. Both models converge toward similar distributions near the end of reionisation. This is because, in the deterministic model, massive haloes ($\mhalo\gtrsim10^{11}\Msun$) dominate the ionising emissivity and generate large, coherent ionised regions. Introducing stochasticity redistributes the emissivity across the halo population and modulates the luminosities of these rare, bright sources, slowing the growth of the largest bubbles and producing a more fragmented ionisation topology.

\begin{figure*}
    \centering
    \includegraphics[width=\textwidth]{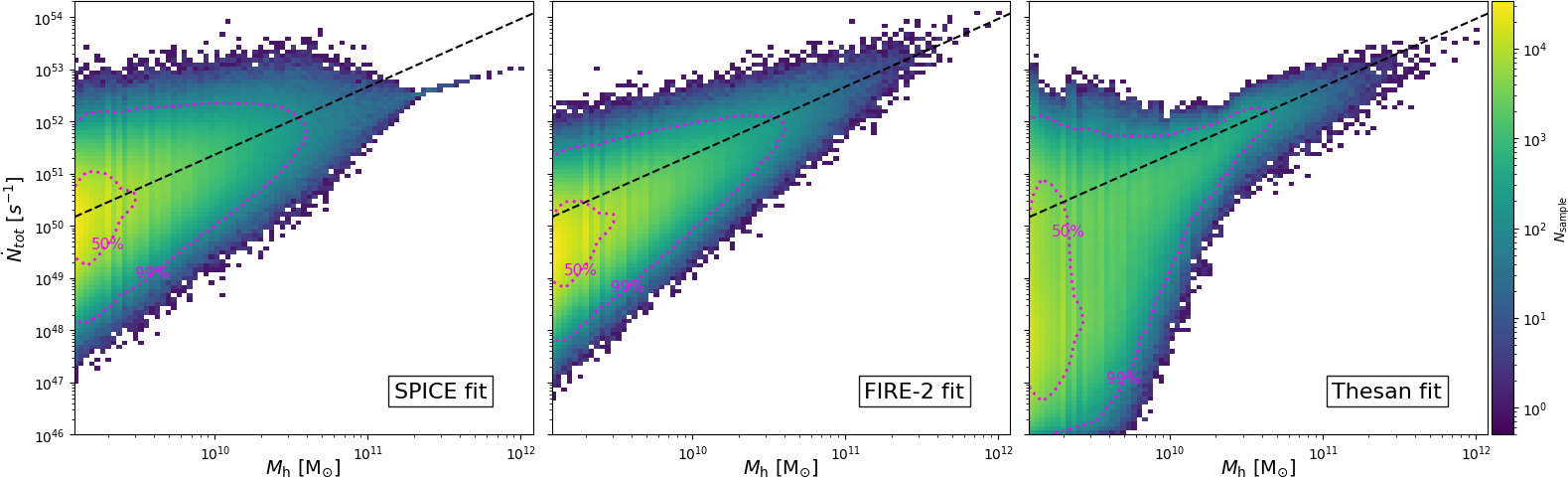}\vskip-2mm
    \caption{Density distribution of the correlation plot $\mhalo-\dotN$ at $z=9$ from the three source models we employed, derived from self-consistent galaxy-formation hydro-simulations during the EoR. The left panel shows the \spice{} bursty model, the middle panel shows the \fire{} model, and the right panel shows the \thesan{} source model. The black dashed line corresponds to our fiducial model for reference. Pink dotted lines indicate the $50\%$ and $90\%$ contours, and a colorbar indicates the sample size.}
    \label{fig:dotN_compare_test2}
\end{figure*}

\subsection{Source Model from Galaxy Formation Simulations}\label{sec:test2}
Previous studies have measured star-formation activity and ionising-photon escape fractions in self-consistent galaxy formation models using hydrodynamical EoR simulations. Here, we consider three such simulations: \fire{} \citep{Sun2023Bursty}, \spice{} with bursty supernova feedback model \citep{Basu2025variabilityuvlf}, and \thesan{} \citep{kannan2022introducing}. We use these simulations to construct physically motivated source models for our emissivity prescription (\autoref{eq:dotN}) and to investigate how different galaxy-formation scenarios affect reionisation.

For the \fire{} based model, we employ their extrapolated analytical star-formation relation \citep[see eq. 4 in][]{Ma2018fire2}, which uses a double power-law that depends on $\mhalo$ and $z$, and convert the resulting quantity into a mean UV magnitude using \autoref{eq:Muv}. We then introduce stochasticity by drawing the UV magnitude from a normal distribution with scatter given by \autoref{eq:std_Muv}. We adopt $a_s=-0.33$ and $b_s=4.5$, consistent with the scatter measured in \fire{} \citep{Gelli2024}, and convert each realisation back into a star-formation rate. For the escape fraction, \cite{Ma2020nomis} measured values between $0.1$ and $0.2$, for haloes with $M_\mathrm{halo}\geq10^9,\Msun$, averaged over multiple halo realisations across a few tens of \fire{} simulations. Therefore, for simplicity, we employ a constant escape fraction of $\fesc=0.1$ and $0.2$ in the \fire{} fit models, as well as an additional extreme case with $\fesc=1$ to explore the maximum ionising efficiency. 

Similarly, for the \spice{} based model, we use tabulated mean and scatter of the star-formation rate provided as a function of halo mass and redshift in \cite{Basu2025variabilityuvlf}. We sample these distributions directly to generate stochastic star-formation histories for our halo catalogue. The \spice{} simulation also constrains LyC escape fractions, showing relatively weak dependence on stellar mass over $M_\star=10^6-10^9\Msun$ and a decrease toward lower redshift. Therefore, we adopt a mass-independent escape fraction of $\fesc=0.068$, matching the global escape fraction measured for the bursty supernova feedback model \citep{Bhagwat2024spice}. 

For the \thesan{} model fit, we use the publicly available source catalogues\footnote{Data are available online: \url{https://thesan-project.com/thesan}} from \cite{Yeh2023thesanfesc}, and model the density distribution of the $\mhalo-\dotN$ relation using kernel density estimation (KDE). For the escape fraction, we include their sub-grid model for photon escape from the stellar birth cloud and ISM with $\fesc^\mathrm{cloud}=0.37$. 

Because all three simulations are limited to volumes smaller than $100\cMpc$ and resolve haloes only up to approximately $10^{10}\Msun$, their mass range is smaller than that covered by our $200\cMpc$ simulation volume. To extend the source models to higher halo masses, we extrapolate the $\mhalo-\dotN$ relation linearly while preserving the scatter measured in the highest resolved mass bin of the corresponding hydrodynamical-based model.

In \autoref{fig:dotN_compare_test2}, we show the resulting halo mass-to-ionising emissivity relation at $z=9$ for the three models. From left to right, we present \spice{}, \fire{} (with $\fesc=0.2$), and \thesan{} fits. Each panel shows the fiducial model for comparison. 

The \spice{} fit predicts ionising emissivity similar to the fiducial model for low-mass haloes, $\mhalo\sim10^9\,\Msun$. Approximately $50\%$ of these haloes have an emissivity of $\dot{N}_\gamma\sim1.5\times10^{50}\,\mathrm{s^{-1}}$. The scatter is broad and evolves slowly for increasing mass, remaining nearly constant, while at higher masses it rapidly decreases toward a one-to-one relationship. This is due to the low statistics from the small number of high-mass halos present in \spice{} at $\mhalo\gtrsim10^{11}\Msun$. As a consequence, the \spice{} fit presents some small-mass haloes that can be more luminous than high-mass haloes. Moreover, at higher mass, the emissivity is about an order of magnitude smaller than in the fiducial model. Since the stochastic model implemented in \pctworay{} uses a generalised version of the \fire{} scatter prescription, the \fire{} fit model exhibits a similar distribution to our stochastic model, \autoref{fig:scatter_model}. However, the emissivity is lower, with approximately $50\%$ of haloes having $\dot{N}_\gamma$ below the fiducial model by $\sim3\times10^{49},\mathrm{s^{-1}}$, due to the choice of free parameters in the observational fit, as discussed in \S\ref{sec:obs_fit}.

Compared with the other two models, the \thesan{} fit scenario produces a broader emissivity distribution at low halo masses, $\mhalo\leq10^{10}\,\Msun$, spanning $\dot{N}\gamma\sim10^{46}-10^{53}\,\mathrm{s^{-1}}$. The original \thesan{} catalogues show a bimodal distribution at low masses, with peaks around $\dot{N}\gamma\sim10^{48}$ and $10^{51}\,\mathrm{s^{-1}}$. However, this feature is absent in our model because our N-body simulation run with \texttt{Pkdgrav3} has limited mass resolution, as discussed in \S\ref{sec:sim_lss}.

\autoref{fig:uvlf} shows the UVLF calculated from the \spice{} fit (orange solid line) and \fire{} fit (red solid line) source model. At $z\simeq 13$, the former case shows a UVLF that matches the faint-end distribution of the stochastic observational fit and the bright-end of the deterministic observational fit. At lower redshift, the \spice{} fit exceeds the expected observation and shows a broad distribution that peaks at $\Muv\simeq-16$, corresponding to an emissivity of $\dotN\simeq7.7\times10^{50}\,\mathrm{s^{-1}}$, which corresponds approximately to the average emissivity of the $99\%$ of the haloes sampled, as we can observe in \autoref{fig:dotN_compare_test2}. We believe this behaviour is connected with a sudden increase in the expected value of the $\sfr$ distribution measured at $z\sim 8$ in the bursty star formation model implemented in the \spice{} simulation.

In contrast, our fit of the \fire{} source model underestimates the UVLF, especially at $z>10$, and agrees reasonably well with the observational data only at lower redshift. This may be because the standard deviation or expected value derived from the $\sfr-\mhalo$ distribution is too small. The discrepancy in the \fire{} fit stems from its limited mass range due to the simulated volume \citep{iliev2014simulating}, as it misses the formation of the largest haloes, with $\mhalo<10^{10}\Msun$ at $z=10$. 

\begin{figure*}
    \centering
    \includegraphics[width=0.83\textwidth]{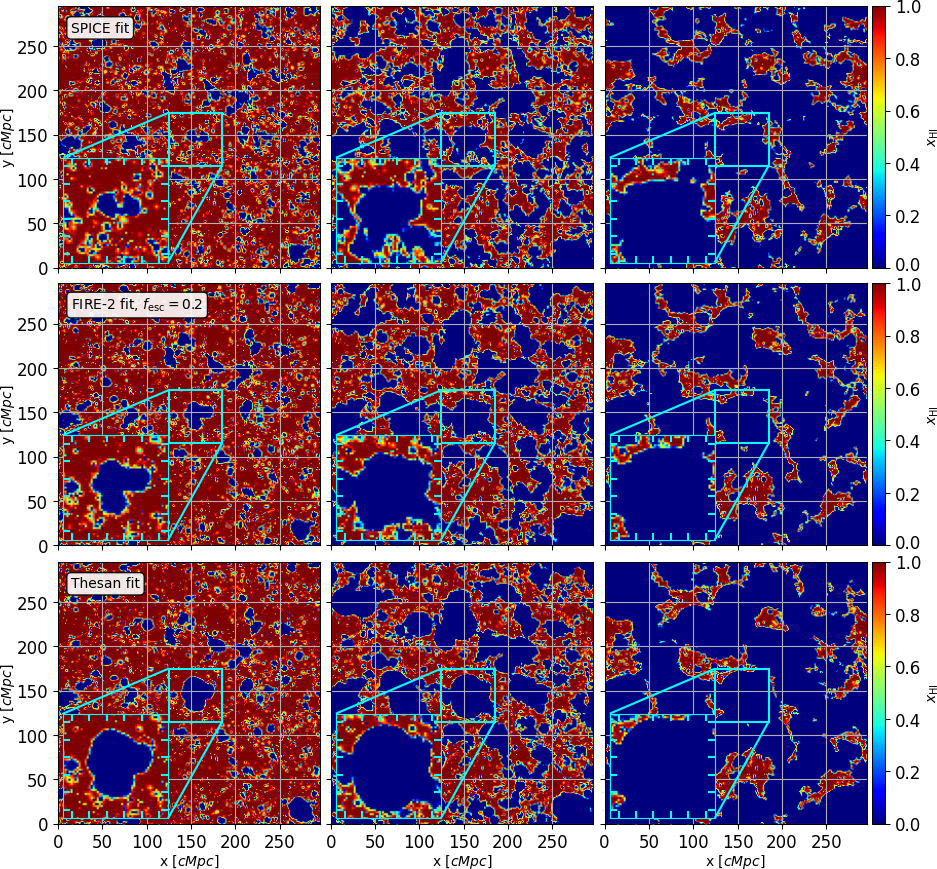}\vskip-2mm
    \caption{Visual comparison of the hydrogen neutral fraction distribution, similar to \autoref{fig:xHI_compare_obsfit}, but for the EoR run with the hydrodynamical-based source model \spice{} (top row), \fire{} (middle row) and \thesan{} (bottom row) at $\overline{x}_\mathrm{HI} = 0.8$ (left column), $0.5$ (middle column) and $0.2$ (right column).}
    \label{fig:xHI_compare_test2}
\end{figure*}

In \autoref{fig:eor_obs_constraint}, we show the resulting reionisation histories using the \texttt{SPICE}, \texttt{FIRE-2}, and \texttt{Thesan} hydro-simulation-derived source models, together with current indirect EoR constraints. The first differences between the models appear in the top-left panel at $z\lesssim14$, where the different source prescriptions begin to produce distinct reionisation histories. The \fire{} model with $\fesc=1$ (red line) evolves fastest, completing reionisation by $z\simeq7.5$. Since this model shares a similar stochastic star-formation prescription with our implemented source model and assumes an extremely efficient UV photon escape fraction, its reionisation history closely resembles the deterministic and stochastic observational fit (observational fit) discussed in \S\ref{sec:obs_fit}. This agreement confirms our previous conclusion that the observationally calibrated models suggest an escape fraction that is likely too efficient. Moreover, this result seems in contrast with the conclusion on the UVLF from \autoref{fig:uvlf}, in fact, the lack of bright sources in the \fire{} fit model could be expected to result in a late reionisation scenario. Instead, because the average $\sfr$ increases rapidly with mass, high-mass haloes drive reionisation. The \spice{} fit (orange line) produces a more rapid transition, beginning around $z\sim12$ and completing by $z\simeq7$. This occurs earlier than the stochastic observational fit and at a similar time as the deterministic observational fit and \fire{} fit with $\fesc=1$. In contrast, \fire{} fit models with $\fesc=0.1$ and $0.2$ (red shaded region) produce a slower reionisation history, consistent with both the fiducial model and the measurement of the final stages of reionisation at $z < 5.7$ \citep{gaikwad2023measuring}. This is the only hydrodynamical-based source model that agrees with these observational constraints on $\overline{x}_\mathrm{HI}$. A similar behaviour appears in the \thesan{} fit model, where the global average neutral fraction decreases gradually for $z<14$ before rapidly declining at $z\approx8$. 

The top-right panel of \autoref{fig:eor_obs_constraint} shows the averaged photoionisation rate derived from our hydro-simulation-based source models, $\langle\Gamma_\mathrm{HI}\rangle$. Meanwhile, the bottom-left panel shows the ionising emissivity per unit volume, $\dot{N}_\mathrm{\gamma,tot}$. Models with efficient ionising sources, including \fire{} fit with $\fesc=1$, \thesan{} fit, and \spice{} fit, generally have a high photoionisation rate and overestimate it at lower redshift. However, these models can reproduce the observed emissivity constraints at $z<6$. This suggests the discrepancy in the averaged photoionisation rate does not originate solely from ionising-photon production but from underestimating the effect of unresolved radiation sinks in the IGM in our simulations \citep{mao2020impact, bianco2021impact, georgiev2024forest}.

The \thesan{} fit model, which shows the strongest redshift evolution in $\overline{\Gamma}_\mathrm{HI}$. At early times, by $z\simeq14$, the characteristic ionised regions in the \thesan{} fit model rapidly approach the imposed mean-free-path scale of $\sim 10\,\mathrm{cMpc}$, as mentioned in \S\ref{sec:sinks}. Once this limit is reached, further increases in the photoionisation rate require either additional ionising sources or a larger $R_\mathrm{mfp}$. The latter occurs only at $z\approx8$, leading to the rapid evolution observed in $\overline{\Gamma}_\mathrm{HI}$. This effect of the mean-free-path also appears, though less pronounced, in the other models. It is characterised by a flattening, followed by a change in the slope of the photoionisation rate, that occurs shortly before the exponential growth of $R_\mathrm{mfp}$ around $z\approx8$. The fluctuations in the \thesan{} fit photoionisation rate mainly reflect the broad scatter in the emissivity of low-mass haloes, as shown in \autoref{fig:dotN_compare_test2}. In \autoref{tab:summary}, we summarise the redshift when reionisation is $\overline{x}_\mathrm{HI}=0.8$, $0.5$ and $0.2$, as well as the Thomson optical depth, for the different runs.

\begin{table}
    \centering
    \caption{Redshift of the reionisation milestones, $\overline{x}_\mathrm{HI} = 0.8,\,0.5$ and $0.2$ and the electron optical depth achieved by the different source models.}
    \begin{tabular}{M{2.5cm}|M{1.0cm}M{1.0cm}M{1.0cm}M{1.0cm}}\hline
       Label                      & $z_{80\%}$ & $z_{50\%}$ & $z_{20\%}$ & $\tau_e$\\ \hline\hline
       fiducial model                & 8.19       & 7.08 & 6.38 & 0.055 \\
       stochastic model                & 8.19       & 7.08 & 6.38 & 0.055 \\
       deterministic obs. fit & 10.64      & 9.30 & 8.41 & 0.079 \\
       stochastic obs. fit  & 9.92       & 8.52 & 7.50 & 0.070 \\ 
       \spice{} fit                & 8.294 & 7.976 & 7.778 & 0.068 \\
       \fire{} fit, $\fesc=1$     & 10.27      & 9.16 & 8.41 & 0.068 \\
       \fire{} fit, $\fesc=0.2$   & 8.29       & 7.24 & 6.58 & 0.056 \\ 
       \fire{} fit, $\fesc=0.1$   & 7.41       & 6.38 & 5.82 & 0.048 \\
       \thesan{} fit                & 8.19       & 7.16 & 6.51 & 0.057 \\ \hline
    \end{tabular}
    \label{tab:summary}
\end{table}

\begin{figure*}
    \centering
    \includegraphics[width=\textwidth]{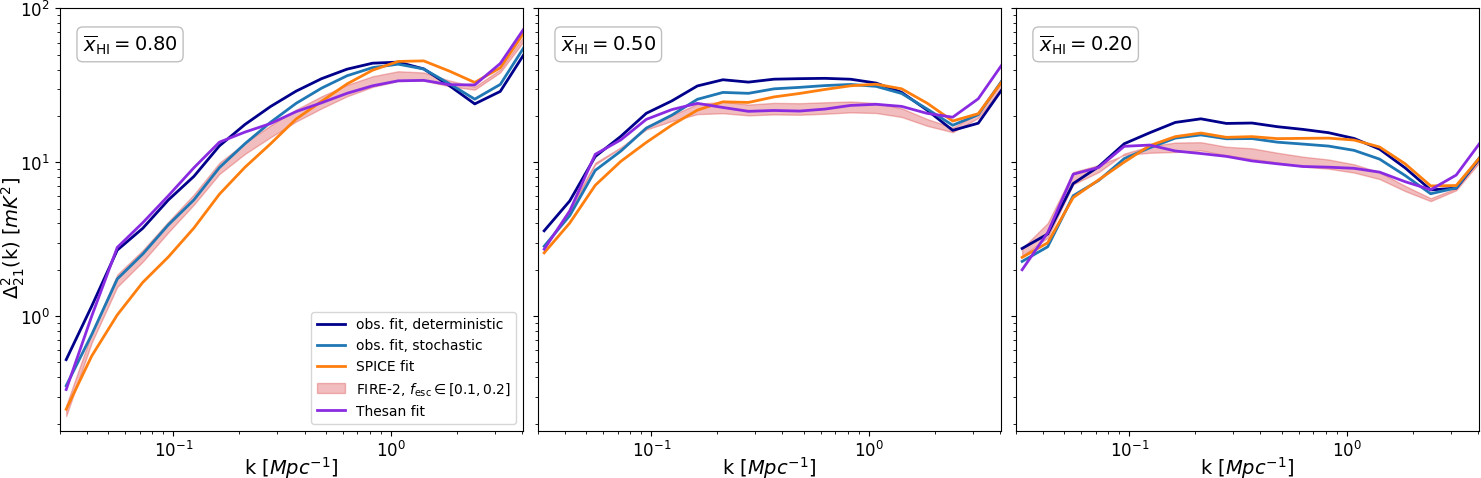}\vskip-2mm
    \caption{Dimensionless power spectra at different EoR milestones, $\overline{x}_\mathrm{HI} = 0.8$ (left column), $0.5$ (middle column) and $0.2$ (right column). Different colours indicate different source model scenarios. We include the result from \autoref{fig:stat_compare_obsfit} (left panel) for comparison.}
    \label{fig:pk_compare_test2}
\end{figure*}

The bottom-right panel shows the integrated Thomson scattering optical depth. We include the current constraint from \cite{collaboration2020planck}, with an observational value of $\tau^\mathrm{Planck}_e = 0.0561 \pm 0.0071$ (grey shaded area). Only \fire{} fit with $\fesc\in[0.1,\,0.2]$ and \thesan{} fit present a value within the $1\sigma$. Interestingly, despite the high $\overline{\Gamma}_\mathrm{HI}$ at high redshift, the \thesan{} fit model remains consistent with the optical depth constraint. This indicates that early ionisation in this scenario is dominated by a small population of highly efficient sources formed in overdense regions, and that their limited abundance regulates the ionisation fraction. The \fire{} fit source model shows that low-mass sources with stochastic star formation can also contribute significantly during the early stages of reionisation, provided the escaping fraction remains limited to $\fesc < 0.2$.

\begin{figure}
    \centering
    \includegraphics[width=\columnwidth]{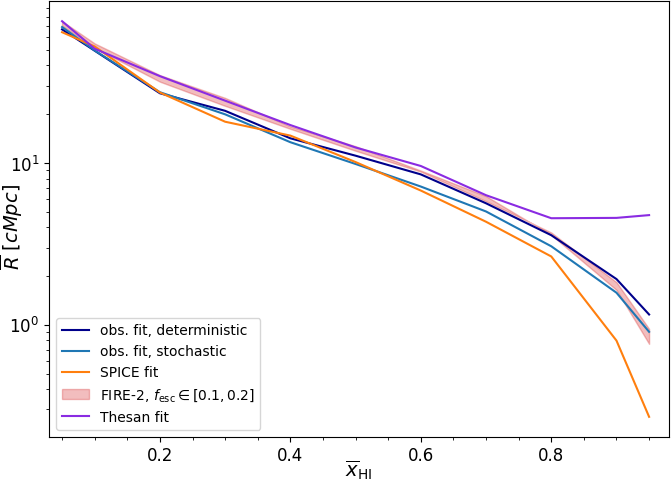}\vskip-2mm
    \caption{Evolution of the mean bubble size, $\overline{R}$, at different stages of reionisation. Different colours indicate different source model scenarios.  We include the result from \autoref{fig:stat_compare_obsfit} (right panel) for comparison.}
    \label{fig:meanR_test2}
\end{figure}

In \autoref{fig:xHI_compare_test2} we provide a visual comparison of the hydrogen neutral fraction distribution at matching EoR milestones, similar to what is presented in \S\ref{sec:obs_fit} for the hydrodynamical-based source models. An inset panel of a $60\,\mathrm{cMpc}$ region zooms in on a large ionised structure containing one of the largest collapsed structures, illustrating the contribution of high-mass sources across the different models. We complement this visual analysis with the statistical comparison of the dimensionless 21-cm power spectra $\Delta^2_\mathrm{21}$ in \autoref{fig:pk_compare_test2}, and the evolution of the mean ionised bubble size $\overline{R}$ in \autoref{fig:meanR_test2}.

Interestingly, the \thesan{} fit model produces ionised regions that are systematically larger and more spherical than the other scenarios. This is confirmed by the BSD analysis, where the mean bubble size reaches $\overline{R}\simeq5\,\mathrm{cMpc}$ at $\overline{x}_\mathrm{HI}=0.9$, compared with $\overline{R}\lesssim1\,\mathrm{cMpc}$ for the other models. This behaviour reflects the contribution of rare massive haloes, which dominate the ionising emissivity and generate large coherent ionised regions.

Similarly, \fire{} fit with $\fesc\in[0.1,\,0.2]$ produces relatively large ionised structures. However, these regions are less spherical and more closely resemble the deterministic observational fit model discussed in \S\ref{sec:obs_fit}. In this case, large \hii{} regions begin to merge, causing the mean bubble size to evolve from values similar to the observational fit towards the larger structures observed in the \thesan{} fit. In contrast, the \spice{} fit produces the smallest ionised regions throughout most of reionisation, reaching comparable bubble sizes only after the formation of the large percolating ionised cluster at $\overline{x}_\mathrm{HI}\sim0.2$. The irregular morphology of this model is mainly driven by the weaker contribution of high-mass haloes in its $\mhalo-\dotN$ relation.

The dimensionless power spectra further demonstrate that different source models produce distinct reionisation morphologies, even when the global neutral fractions match. At large scales, $k=0.1\,\mathrm{cMpc}^{-1}$, the models exhibit significant differences already at $\overline{x}_\mathrm{HI}=0.8$, with \thesan{} fit and deterministic observational fit reaching $\Delta^2_\mathrm{21}\approx6\,\mathrm{mK^2}$, \fire{} fit and stochastic observational fit reaching $\Delta^2_\mathrm{21}\approx4\,\mathrm{mK^2}$, and \spice{} fit remaining near $\Delta^2_\mathrm{21}\approx2\,\mathrm{mK^2}$.

The rapid evolution between $k=0.1\,\mathrm{cMpc}^{-1}$ and $1\,\mathrm{cMpc}^{-1}$ reflects the range of ionised bubble sizes generated by the different source prescriptions. These differences are largest at early times ($\overline{x}_\mathrm{HI}=0.8$) and are driven primarily by high-mass halo efficiency and these haloes ability to form large ionised regions quickly. In particular, the power-spectrum turnover scale estimates the characteristic bubble size. For the \thesan{} fit model, a rapid decrease in power occurs at $k\sim0.2\,\mathrm{cMpc}^{-1}$, whereas the other models show a smoother transition that starts at $k\sim1\,\mathrm{cMpc}^{-1}$, consistent with their corresponding $\overline{R}_\mathrm{mfp}$ of $\sim 5\cMpc$ and $\sim 1.25\cMpc$, respectively.

At later stages, the differences between the models decrease. At $\overline{x}_\mathrm{HI}=0.5$, all models reach comparable large-scale amplitudes of approximately $10-20\,\mathrm{mK^2}$, while by $\overline{x}_\mathrm{HI}=0.2$ the differences reduce further to around $10\,\mathrm{mK^2}$. The \spice{} fit model maintains a more rapidly evolving power spectrum, reflecting the continued growth of smaller ionised structures before percolation. On small scales, $k>1\,\mathrm{cMpc}^{-1}$, all models converge to similar amplitudes with values that vary around $k\sim10\,\mathrm{mK^2}$, indicating that the largest differences between source models are encoded primarily in the large-scale topology.


\section{Conclusions}\label{sec:conclusion}
In this work, we implemented a stochastic source model in the \pctworay{} code for EoR simulations. This created a unified framework that allows direct comparison between source models derived from hydrodynamical simulations of galaxy formation and models parametrised by observational constraints, to quantify the impact of stochastic star formation on the reionisation process. To do so, we used an N-body simulation of volume $\left(295\cMpc\right)^3$, with a minimum halo mass of $\sim 10^9\Msun$, and post-processed it with \pctworay{} to obtain the $\hi{}$ distribution between redshift $20$ and $5$. We assume that ionising sources in lower halo masses are fully suppressed by radiative feedback.

We developed an ionising photon source model in \pctworay{} that maps to absolute UV magnitude and a stochastic star formation model, based on the scatter measured in hydrodynamical simulations, which we can parametrise with two free parameters. We sample stochasticity in the $\Muv-\mhalo$ relation from a Gaussian distribution that propagates to the $\sfr$. Moreover, we improved the UV escape fraction, modelled as a double power law in redshift and stellar mass, to better capture its evolution.

At first, we compare results from two EoR runs with the same total photon production, $\dot{N}_\mathrm{\gamma,tot}$. The former uses stochastic star formation (stochastic model), while the latter uses our fiducial model with a deterministic relation (fiducial model). Our first conclusion is that despite the statistical difference between the two models, if the maximum relative difference in the photon production is within $|\Delta \dotN^\mathrm{tot}| \leq 2 \%$ throughout reionisation, then the relative deviation in the global neutral fraction is limited to $|\Delta x_\mathrm{HI}| \sim 10^{-3}$. Consequently, differences in the 21-cm summary statistics are imperceptible and insufficient to uniquely determine the statistical nature of the $\sfr-\mhalo$ relation. In contrast, \hii{} map analysis reveals small differences around ionised regions. These stem from bursty star formation at high redshift that leaves an imprint on the ionisation front, making the hosting \hii{} bubble appear larger and remain visible at later times, long after the burst event ended.

We then calibrate the free parameters of the deterministic and stochastic setups (observational fits) using Bayesian inference on JWST high-redshift galaxy observations, and motivate the need for scatter in the star formation rate to interpret the UVLF at high redshift. We constrain the free parameters of the stellar-to-halo ratio, $f_\mathrm{\star,0}$, $\alpha_\star$, the accretion model, $\alpha_\mathrm{h}$, the scatter, $a_s$ and $b_s$ from JWST measurements of the UV luminosity function. For the free parameters of the UV escape fraction, $f_\mathrm{esc,0}$, $\alpha_\mathrm{esc,M}$ and $\alpha_\mathrm{esc,Z}$, we employ observational data of the Lyman-continuum leakage.


Posterior analysis suggests a stronger evolution of the escape fraction over cosmic time than previously assumed, indicating that it should decrease with increasing stellar mass and decreasing redshift. Both observational fit scenarios yield similar values of the accretion exponential model, $\alpha_\mathrm{h}$, larger than in previous studies of dark matter halo formation. In the context of reionisation, this suggests that accretion of fresh neutral gas to fuel new star formation is substantially attenuated. We find that the deterministic observational fit deviates from the observed UVLF at progressively higher redshift, showing a similar trend to our fiducial model. Meanwhile, as in previous work, our stochastic observational fit flattens the UVLF by increasing the contribution of bright, low-mass haloes ($M_\mathrm{halo}\lesssim10^{10}\Msun$) to the ionising photon budget, suggesting that these systems play a more prominent role in driving reionisation. 

Moreover, the deterministic observational fit shows faster reionisation, mostly due to a larger amplitude, $f_\mathrm{\star,0}$. We analysed the resulting dimensionless power spectra at matching reionisation milestones: $\overline{x}_\mathrm{HI}=0.8$, $0.5$, and $0.2$. A consistently positive bias at large scales in the deterministic scenario ($k\leq1\,\mathrm{Mpc^{-1}}$) indicates a more spatially clustered ionisation field, in which massive haloes drive the growth of large, coherent ionised regions while leaving vast regions of the IGM relatively neutral. In contrast, stochastic star formation redistributes ionising emissivity toward lower-mass haloes, resulting in a more fragmented ionisation topology with smaller bubbles and reduced large-scale contrast. These conclusions are confirmed by \hi{} maps and bubble size distribution (BSD) analysis, which show that the deterministic observational fit model consistently produces larger ionised regions at $\overline{x}_\mathrm{HI}\geq0.4$.

For our third result, we employ source models derived from hydrodynamical simulations of galaxy formation during EoR. In our case, we consider the \spice{}, \fire{} and \thesan{} simulations. We employ fitting tables and equations for $\sfr$ and the associated scatter, or, when provided, the total emissivity directly. We first note that the inferred UVLF from these models does not match observational evidence across redshifts. In the \spice{} fit, the model first underestimates the UVLF at $z>11$ and then overestimates it at $z<8$, with a broad distribution peaking at $\Muv\simeq-16$. Meanwhile, the \fire{} fit underestimates the bright end at higher redshift. We conclude that these limitations reflect the mass limit imposed by the choice of the simulated volume. Hydrodynamical simulations generally focus on small cosmological volumes ($\leq100\cMpc$) to resolve the virial mass of atomic-cooling haloes, but this limits the high-mass end of the halo mass function. Therefore, we suspect that to match the observed UVLF, these simulations rely too heavily on the abundance of low-mass haloes to compensate for the missing large haloes. As a consequence, they require a less extreme stochasticity in the star-formation rate. In contrast, as in our case, selecting a large volume to account for sampler variance in the ionised regions requires a stronger scatter, since we assume that $\leq10^9\Msun$ are suppressed and do not contribute to the UVLF.

We then run \pctworay{} with the hydrodynamical-based source models mentioned above and show the resulting tomographic maps from the different runs, showing that the different models yield distinguishable maps in morphology and \hii{} bubble size distribution at matching reionisation milestones. We show indirect EoR observables: reionisation history, average photoionisation rate, ionising photon emission rate per unit volume and the electron optical depth.  

The source model based on hydrodynamical simulations agrees well with the overall reionisation history. However, they tend to underestimate the global average neutral fraction by one order of magnitude during the final stage of EoR. The \fire{} fit is an exception: if we make a conservative assumption on the escape fraction, $\fesc\in[0.1,0.2]$, then it remains in reasonable agreement with EoR observational data. Moreover, none of our reionisation scenarios, with the presented hydrodynamical-based source models and observational fit, can match all four reionisation constraints concurrently, indicating that current reionisation simulation frameworks miss relevant reionisation physics. This could be due to the high-resolution hydrodynamical EoR simulation approach, our choice and implementation of the source and sink model, or missing relevant sub-grid physics in both high-resolution and large-scale simulations. Notable differences between the source models appear in 21-cm summary statistics at matching milestones, such as dimensionless power spectra and bubble size statistics. 

This work demonstrates that the spatial distribution of ionising sources, together with the coupling between star-formation efficiency, ionising escape fraction, and IGM absorption, plays a fundamental role in determining the timing and morphology of cosmic reionisation. Although different source models can reproduce similar global reionisation histories, they generate distinct ionisation topologies and characteristic 21-cm statistical quantities. Stochastic star formation redistributes ionising emissivity towards lower-mass haloes, producing a more fragmented ionisation structure while delaying the growth of the largest ionised regions.

Source models calibrated from different hydrodynamical simulations likewise predict diverse ionising emissivity distributions, leading to measurable differences in the evolution of \hii{} bubble size distribution and the 21-cm power spectrum. Our results underscore the importance of developing a unified framework between galaxy formation and EoR large-scale simulations to address the multi-scale challenge the reionisation process poses. Moreover, our results highlight the importance of combining forthcoming large-scale 21-cm statistics with JWST measurements of the UV luminosity function and Lyman-continuum leakage, which provide direct sensitivity to the reionisation topology and allow us to distinguish competing source models.

\section*{Acknowledgements}
The authors would like to thank Pascal Oesch and Emma Giovinazzo for sharing their data. MB acknowledges the financial support from the ETHZ Cost Centre, grant number 23928. SKG acknowledges support from the Olle Engkvist Stiftelse (grant no. 232-0238). EG is supported by the JSPS KAKENHI grant ILR 23K20035. TB was supported by the Platform for Advanced Scientific Computing (PASC) project \textit{"Accelerated Radiative Transfer Simulations for SKAO"} (2024-29). We acknowledge the Nordita program called \textit{Cosmic Dawn at High Latitude}, which gave us the opportunity to discuss and initiate this project. We acknowledge access to Piz Daint at the Swiss National Supercomputing Centre, Switzerland, under the SKA share with project IDs \texttt{sk014} and \texttt{c45} from PASC.

\section*{Data Availability}
The data underlying this article are available upon request and can also be regenerated from scratch using the publicly available \texttt{Pkdgrav3} \citep{potter2017pkdgrav3}, \pctworay{} \citep{Hirling2024pyc2ray} and \texttt{tools21cm} \citep{giri2020tools21cm}. Tables from the \spice{} simulations are available upon request by the author of \cite{Basu2025variabilityuvlf}. Observational data from JWST and reionisation constraints are from \cite{Garaldi2023corecon}.

\bibliographystyle{mnras}
\bibliography{bibliography.bib} 

\appendix

\bsp	
\label{lastpage}
\end{document}